\documentclass[preprint,journal]{vgtc}            % preprint (journal style)

\newtoggle{show_comments}
\togglefalse{show_comments}

\newtoggle{anonymous}
\toggletrue{anonymous}
\newtoggle{highlight_revisions}
\toggletrue{highlight_revisions} 
\newcommand{\new}[1]{\textcolor{black}{\iftoggle{show_comments}{new: #1}{#1}}}
\newcommand{\add}[1]{\textcolor{black}{\iftoggle{show_comments}{new: #1}{#1}}}

\onlineid{0}

\vgtccategory{Research}

\title{ResonaVis: Visualizing Interactive Music Data to Support Reflective Music Composition for Therapeutic Contexts}

\author{%
  \authororcid{Abhishek Karwankar}{0009-0005-9878-6519},
  \authororcid{Elise Ruggiero}{0009-0003-2617-7138},
  \authororcid{Daniel Stevens}{0009-0000-5646-0282},
  \authororcid{Matthew Louis Mauriello}{0000-0001-5359-6520}
}

\authorfooter{
  \item
  	All authors were affiliated with the University of Delaware at the time of work. E.R. is now affiliated with the Florida State University.
  \item 
    Emails: karwabhi@udel.edu, emrugg@udel.edu, stevens@udel.edu, mlm@udel.edu 
}

\abstract{\new{Designing music for therapeutic contexts requires navigating complex relationships between musical structure and listeners' sensory responses, yet composers often lack structured representations of these interactions \new{to inform compositional decisions}, relying instead on intuition. We present ResonaVis, an interactive visualization system to \new{help composers analyze interaction and audio data from prior sessions with} children with Autism Spectrum Disorder (ASD)\new{, informing future compositions}. ResonaVis integrates audio features and interaction logs to capture how children engage with layered musical compositions, representing this engagement through coordinated visualizations of temporal transitions, layer co-occurrence, rhythmic activity, and spectral characteristics. \new{Rather than prescribing strategies or directly supporting therapy sessions}, the system surfaces patterns \new{in past} session data, supporting data-informed reflection during iterative composition. We evaluated ResonaVis through a mixed-methods study with eight music students \new{and} a follow-up case study with two experienced composers. Results demonstrate good usability (SUS = 72.23), exceeding benchmarks for early prototypes, and indicate that participants were able to identify interaction patterns, reason about layer relationships, and make informed compositional decisions. Despite moderate cognitive demands, participants reported high perceived performance and low frustration, suggesting productive engagement. Participants' confidence \new{in interpreting interaction and acoustic data} when composing for individuals with ASD also increased significantly across multiple dimensions—rhythm and dynamics, pitch and timbre, and spectral characteristics (all $p < 0.05$). Finally, qualitative findings suggest the potential to shift composers from intuition-driven toward more adaptive, data-informed composition. This work contributes a visualization design space for therapeutic music interaction data, an integrated system for compositional reflection, and empirical evidence that visualization tools can support analytical reasoning and confidence in data-informed creative practices.}}


\teaser{
  \centering
  \includegraphics[width=\linewidth, alt={Title Image}]{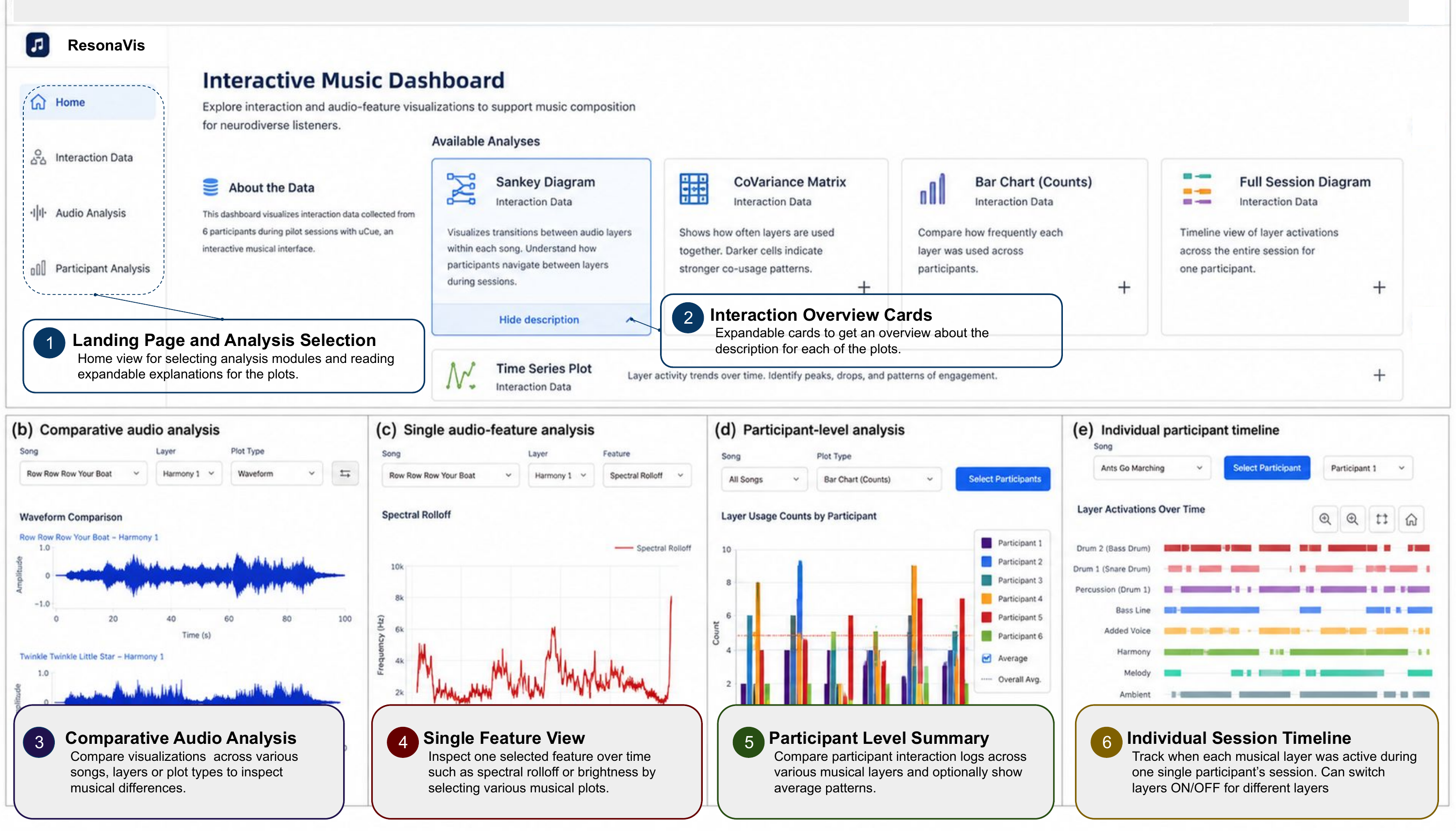}
    \caption{ResonaVis combines interaction-data and audio-feature visualizations to help composers explore participant engagement with musical layers, including analysis selection, comparative audio inspection, participant-level summaries, and session timelines.}
  
  \label{fig:teaser}
}

\graphicspath{{figs/}{figures/}{pictures/}{images/}{./}} % where to search for the images

\usepackage{tabu}                      % only used for the table example
\usepackage{booktabs}                  % only used for the table example
\usepackage{lipsum}                    % used to generate placeholder text
\usepackage{mwe}                       % used to generate placeholder figures
\usepackage{ccicons}                   % package to be able to use icons from creative commons
\usepackage{soul}
\usepackage{enumitem}
\usepackage{textcomp}
\usepackage[utf8]{inputenc}
\usepackage{csquotes}%% <-- added
\usepackage{graphicx} 
\usepackage{subcaption}  % Include the subcaption package
\usepackage{tcolorbox}
\tcbuselibrary{skins}
\usepackage{lipsum}  % For generating dummy text
\usepackage{multirow}
\usepackage[table,xcdraw]{xcolor}
\usepackage{adjustbox}
\usepackage{pifont}
\usepackage{pdflscape}
\usepackage{tabularx}
\usepackage{subcaption}
\usepackage{float}
\usepackage{placeins}
\usepackage{subcaption}
\usepackage{dblfloatfix}
\usepackage{needspace}

\usepackage{mathptmx}                  % use matching math font

\begin{document}

%%%%%%%%%%%%%%%%%%%%%%%%%%%%%%%%%%%%%%%%%%%%%%%%%%%%%%%%%%%%%%%%
%%%%%%%%%%%%%%%%%%%%%% START OF THE PAPER %%%%%%%%%%%%%%%%%%%%%%
%%%%%%%%%%%%%%%%%%%%%%%%%%%%%%%%%%%%%%%%%%%%%%%%%%%%%%%%%%%%%%%%


\maketitle

\section{Introduction}

Music therapy is a widely recognized intervention for children with Autism Spectrum Disorder (ASD), supporting emotional regulation \cite{geretsegger2014music}, communication \cite{gold2004effects}, and social engagement \cite{reschke2011history}. However, some therapeutic approaches rely on passive listening or facilitator-guided participation, limiting opportunities to adapt music to individual sensory preferences and behavioral needs. Because children with ASD exhibit highly individualized responses to rhythm, timbre, and musical structure, therapeutic outcomes depend on how effectively music can be personalized \cite{daniel2022rhythmic, khyzhna2020music}. While prior work has focused on the benefits of music exposure, less attention has been given to how data generated during interactive sessions can inform compositional decisions.

Interactive music systems such as uCue \cite{10.1145/3713043.3727053} enable children to explore music by activating and modifying instrumental layers, producing rich temporal interaction logs that capture when musical elements are added, sustained, or removed. These logs encode behavioral responses to musical structure and offer a potential foundation for data-driven composition. However, they are high-dimensional, multi-layered, and temporally complex, making them difficult to interpret through raw data or listening alone. Composers must reason about patterns such as layer co-occurrence, temporal evolution, and relationships between interaction behavior and acoustic features—tasks that are challenging without appropriate visual representations. Prior work has shown that visualizing interaction data can reveal meaningful patterns and support analytical reasoning that would otherwise remain hidden \cite{10453452}.

Interactive visualization systems, particularly visual analytics dashboards, are well-suited to support such tasks by enabling users to explore heterogeneous data, identify relationships, and derive insights through coordinated views \cite{sarikaya2018we, keim2008visual, 10.5555/832277.834354}. While dashboards have been widely adopted in domains such as healthcare, education, and business intelligence, their role in supporting creative practices—especially in music composition and therapeutic contexts—remains underexplored. This gap motivates an open question: \textit{can interactive visualization support composers in interpreting complex interaction logs to inform therapeutic music composition?} To address this, we design and evaluate \textit{ResonaVis}, an exploratory visual analytics system  that investigates how visual representations can support compositional reasoning.

 ResonaVis integrates multiple coordinated visualizations—including temporal session timelines, layer transition diagrams, co-occurrence matrices, and acoustic feature plots—to enable exploration of interaction patterns across musical layers and \new{listening} sessions. The system was developed through co-design sessions and iterations with composers experienced in therapeutic music composition, ensuring alignment with real-world analytical and creative tasks. We investigate the following research questions: \textit{(RQ1) How can interactive visualization techniques support the process of composing music for children with ASD?} \textit{(RQ2) In what ways can a visual dashboard help composers identify and interpret relevant musical patterns from session data?} And,\textit{ (RQ3) what design implications emerge from music composers engaging with visualization dashboards for therapeutic music composition?}

We evaluate ResonaVis through a mixed-methods exploratory study with eight music students trained in music composition/education and familiar with music therapy, alongside an observational case study with two experienced composers. Participants used ResonaVis to analyze uCue session data and reported significant increase in confidence in composing therapeutic music, particularly in musical aspects such as rhythm and dynamics ($p = 0.0004$), pitch and timbre ($p = 0.038$), spectral energy and brightness ($p = 0.004$), as well as interpretation of visualizations to support music compositional decisions ($p = 0.032$). The system supported the identification of interaction patterns not easily perceived through listening alone and achieved a System Usability Score (SUS) of 72.23 over the two studies, indicating good usability.

This work contributes: \textit{i)} \textit{ResonaVis}, an interactive visual analytics system that integrates temporal interaction logs and acoustic features to support exploratory analysis of therapeutic music sessions. \textit{ii)} Visualization design studies demonstrating how composers interpret complex interaction data to inform creative decision-making. \textit{iii)} Empirical evidence showing how visualization supports insight generation and confidence in therapeutic music composition. \textit{iv)} Design implications for visual analytics systems supporting creative practices in music and other domains involving complex temporal interaction data. 

\section{Related Works}

This work builds on research in music therapy for ASD, interactive and technology-supported music composition, and visual analytics systems for creative sensemaking. \new{We situate our contribution at the intersection of data visualization and music composition to empower composers to interpret listener interaction data and inform therapeutic music design.} 
% We situate our contribution at the intersection of how visualization can support composers in interpreting complex interaction data to inform therapeutic music design.

\subsection{Music Therapy and Interactive Music Systems for ASD}

Music therapy is a well-established intervention for children with ASD, supporting communication, social interaction, and emotional regulation \cite{wigram2006music, whipple2004music, bharathi2019music, lagasse2017social, chen2024music}. Prior work demonstrates that music provides a non-verbal medium for expression, enabling meaningful behavioral and emotional engagement \cite{lagasse2014effects, geretsegger2014music, katagiri2009effect}. However, a growing body of research highlights that therapeutic effectiveness depends on specific musical elements, including rhythm, pitch, timbre, and frequency, which shape sensory and emotional responses. Studies show that individuals with ASD often exhibit distinct sensitivities to these attributes, such as preferences for rhythmic predictability, lower brightness, and stable pitch ranges \cite{allen2009hath, santos2024assessing, dotch2023understanding, michel2024sounds}. Similarly, observational and interview-based studies indicate that lower-frequency sounds and gentle timbral qualities can promote engagement and reduce sensory overload \cite{10.1145/3240925.3240958}. 

While this literature provides important insights into ASD-specific musical preferences, most work focuses on therapeutic outcomes rather than supporting the compositional process itself. In practice, composers must translate these principles into design decisions, often without tools that connect behavioral interaction data with musical structure. Our work builds on these findings by enabling composers to interpret interaction patterns and acoustic features through visualization, supporting alignment between compositional and therapeutic considerations.

\subsection{Music Technology and Creative Confidence}

Music composition is inherently a creative and exploratory process \cite{henry1996creative}. Advances in digital tools and interactive systems have expanded the design space for composers, enabling new forms of experimentation, iteration, and personalization \cite{born2005musical, de2012creativity, smith2009practice}. Technologies such as digital musical instruments and interactive interfaces support adaptive and participatory music-making, particularly in contexts involving children or users with diverse needs \cite{mayer2021music, bjork2005games}. These systems facilitate rapid exploration of musical ideas and enable compositions to evolve dynamically based on user interaction \cite{collins2010introduction, paine2009towards}. Beyond technological affordances, confidence plays a critical role in shaping compositional practice. Prior work shows that higher confidence is associated with increased creativity, willingness to experiment, and the development of distinctive musical styles \cite{sawyer2024explaining, hargreaves2018musical}. In therapeutic contexts, confidence is particularly important, as composers must balance artistic intent with sensitivity to user needs and therapeutic goals \cite{mcferran2014community}. Nonetheless, existing tools primarily support music creation rather than reflective reasoning about how compositions align with user behavior or therapeutic outcomes. This gap suggests an opportunity for tools that not only enable composition but also support interpretation and decision-making. We explore how visual analytics can enhance composers’ confidence by making complex interaction and acoustic data more interpretable.

\subsection{Visual Analytics for Creative Sensemaking}

Interactive visualization systems are widely used to support sensemaking in complex data environments by enabling users to explore patterns, relationships, and trends through coordinated views \cite{keim2008visual, 10.5555/832277.834354, sarikaya2018we}. These systems support iterative exploration and abstraction across data and task levels, aligning with established models of visualization design and analysis \cite{munzner2025visualization}. Visual analytics research emphasizes the role of interactive systems in supporting analytical reasoning, insight generation, and decision-making \cite{pirolli2005sensemaking, north2006toward, chang2009defining}. Dashboards, in particular, have been adopted across domains such as healthcare and education to support monitoring, reflection, and intervention \cite{bernard2015visual, monroe2013temporal, mauriello2016simplifying, naranjo2019visual}. In creative and therapeutic contexts, visualization has been used to represent emotional states, engagement patterns, and creative processes over time \cite{franklin2017dashboard, cybulski2015creative}. However, most systems are designed for retrospective analysis rather than supporting ongoing creative decision-making. Within music-related applications, visualization has been used to analyze audio features and listener responses, such as mapping spectral, rhythmic, or perceptual attributes to emotional interpretations \cite{mcadams2013musical}. Prior work also explores visualizing performance and interaction traces, including gestural input, timing, and improvisational structure \cite{foote1999visualizing, astrid2017moment, fiebrink2011human, xambo2024human}. While these approaches reveal meaningful patterns, they rarely connect interaction data back to compositional workflows. In therapeutic settings, this limitation is particularly significant, as interaction patterns may serve as proxies for user preferences and sensitivities \cite{wigram2004improvisation, kim2008effects}. Our work addresses this gap by framing visualization as a creativity support tool that enables composers to interpret interaction data and translate insights into compositional decisions.

\subsection{Music Visualization and Multimodal Representations}

Visualization techniques for music have traditionally focused on representing individual musical attributes such as waveform, pitch, and spectral structure \cite{hiraga2002performance, fonteles2013creating, lima2021survey}. These representations are widely used in analysis and education but often operate on isolated dimensions of musical data. Similarly, compositional support tools have explored visual representations of rhythm and harmony but typically within constrained or pedagogical settings \cite{10.1145/964696.964698, 10.1145/964696.964700, chan2009visualizing}. Recent surveys highlight that most music visualization systems do not integrate multiple data modalities, such as combining acoustic features with interaction or behavioral data \cite{lima2021survey}. Emerging work has begun to explore visualization as a component of creative workflows, but these systems remain limited in scope and rarely address real-world therapeutic contexts \cite{miller2022augmenting, de2017understanding}. In contrast to traditional digital audio workstations (DAWs), which focus on direct music creation, our approach emphasizes reflective analysis through multimodal visualization. ResonaVis integrates temporal interaction logs with acoustic features such as spectral energy, pitch, and dynamics, enabling composers to explore how user behavior and musical structure interact over time. This multimodal perspective supports a deeper understanding of engagement patterns and provides a foundation for data-informed compositional reasoning. ResonaVis operationalizes this perspective by enabling exploratory analysis of complex session data to inform creative decision-making.

% However, no existing system integrates interaction data, acoustic features, and visual analytics to support composers in therapeutic music design—a gap ResonaVis addresses by enabling exploratory analysis of session data to inform creative decision-making.
% However, there remains a lack of systems that integrate interaction data, acoustic features, and visual analytics to support composers in therapeutic music design. ResonaVis addresses this gap by enabling exploratory analysis of complex session data to inform creative decision-making.

% \smallskip
% Overall, prior work establishes the importance of music therapy, advances in interactive composition technologies, and the potential of visualization for sensemaking. 
% However, there remains a lack of systems that integrate interaction data, acoustic features, and visual analytics to support composers in therapeutic music design. 

\section{Iterative Design Process}
We describe the dataset, design process, composition challenges, dashboard design, and how the visualizations address these challenges.

\subsection{Data and Design Context}

The dataset used in ResonaVis consists of approximately four hours of music interaction logs collected from six children with ASD during guided play sessions using the uCue interface \cite{10.1145/3713043.3727053, mauriello2025interactive}. While derived from sessions with children with ASD, they were not involved in the system’s design or evaluation; instead, the data serve as a proxy for understanding interaction patterns. During these sessions, children interacted with modular musical compositions by activating and deactivating instrumental layers in real time through a tangible button-based interface. Each interaction event corresponding to a button press or release was automatically recorded by the system with timestamps and layer identifiers, producing a temporally ordered log of user actions. Logs were exported as CSVs and are referred to as \textit{Interaction Data}.

The interaction logs are meaningful because they provide a fine-grained, behavioral trace of how children engage with different musical elements over time. Unlike post-hoc observations or subjective reports, these logs directly capture moment-to-moment interaction patterns, including which layers are preferred based on how frequently they are activated and how long they are sustained. As such, they offer an objective basis for understanding how musical structure influences engagement in therapeutic contexts.

Importantly, these logs also serve as a proxy for engagement behavior. Sustained activation of layers can indicate one's continued interest or comfort with specific sounds, while frequent toggling or rapid transitions may reflect exploration, curiosity, or sensitivity to certain auditory stimuli. Patterns such as repeated co-activation of layers, temporal sequencing of musical elements, and duration of interaction provide insights into how children respond to rhythm, timbre, and musical complexity. However, because these behaviors are encoded as high-dimensional temporal data across multiple layers, they are difficult to interpret without appropriate visual representations.

\begin{figure}
    \centering
    \includegraphics[width=1\columnwidth]{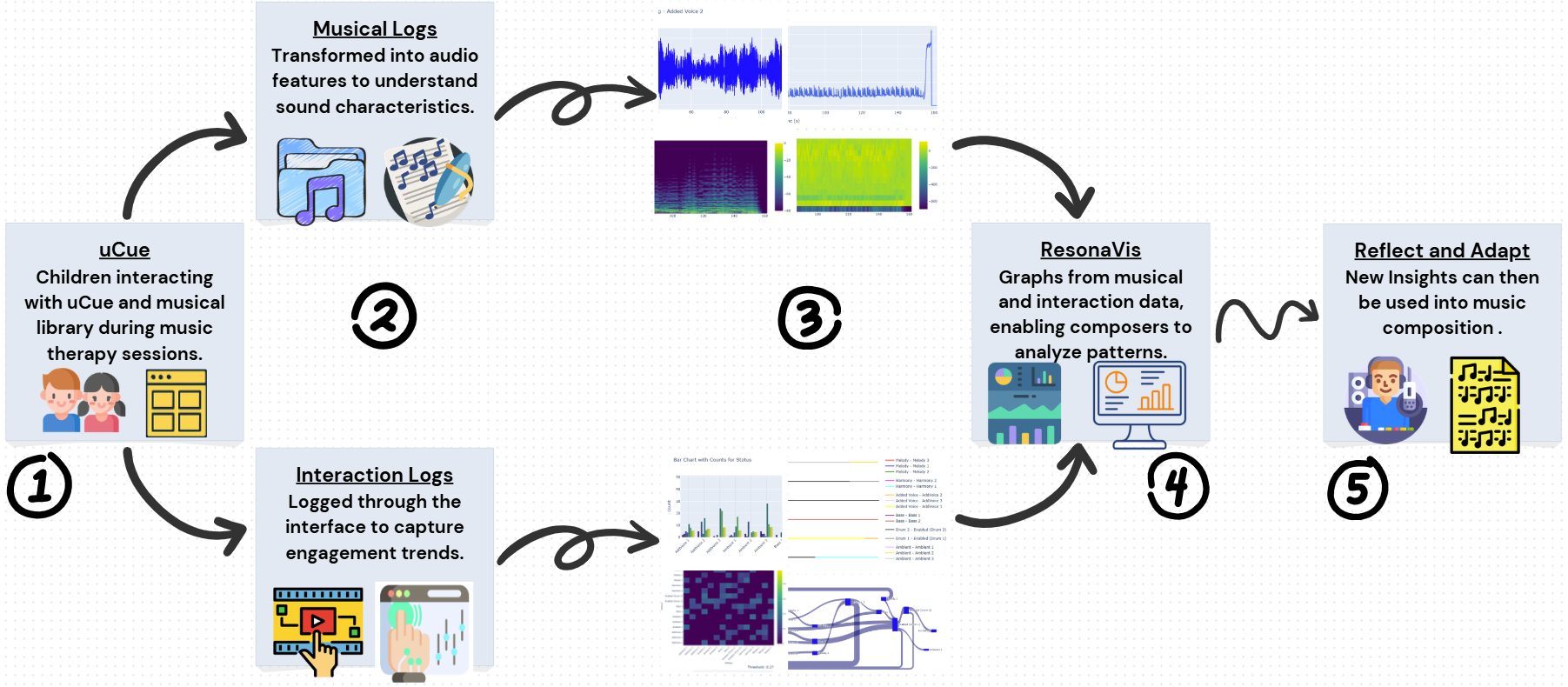}
\caption{\textbf{ResonaVis workflow for data-driven music composition.} (1) Children interact with the uCue system during therapy sessions, generating interaction logs. (2) These logs are processed into interaction-based visualizations. (3) In parallel, musical data is transformed into acoustic feature visualizations. (4) ResonaVis integrates both interaction and musical representation. (5) Insights from this analysis inform the composition of new therapeutic music.}
    \label{fig:combined}
\end{figure}

In addition to interaction data, we incorporate the full collection of musical tracks used in uCue to extract audio features, referred to as \textit{Musical Data}. Together, these two data sources inform \new{the selection of} ten visualizations: five derived from interaction behavior and five from acoustic characteristics. Figure~\ref{fig:combined} illustrates the ResonaVis workflow, showing how interaction and musical data are transformed into coordinated visual representations to support compositional analysis.

% \begin{figure*}[t]
%     \centering
%     \includegraphics[width=0.75\textwidth]{figs/WorkFlow/MusicVIS Diagrams (2).pdf}
%     \caption{\textbf{The MusicVis workflow for supporting therapeutic music composition. The system combines musical and interaction data visualizations. Together, these plots allow composers to analyze how children interact with modular music layers and how the musical features evolve, supporting evidence-based decisions for creating or adapting new therapeutic compositions.}}
%     \label{fig:combined}
% \end{figure*}

\subsection{Participatory Design}

% Participatory Design emphasizes collaboration between stakeholders and designers to create more effective, user-centered solutions \cite{greene1987stakeholder, muller1993participatory, ehn1988work}. We adopted a participatory design approach to develop ResonaVis, collaborating with music composers to ensure that the system reflects real-world compositional reasoning practices. Beyond interface design, our goal was to identify appropriate \textit{visual representations} and \textit{analytical workflows} that support interpreting interaction data generated during therapeutic music sessions. This work follows a visualization design study methodology, iteratively refining data abstractions, task requirements, and visual encodings in collaboration with domain experts \cite{11262783}. Prior work in music and affective technology has similarly demonstrated the value of participatory approaches for designing systems that align with users’ perceptual, creative, and therapeutic needs \cite{van2015participatory, wilkie2013towards, grond2019participatory}.
Participatory Design emphasizes collaboration between stakeholders and designers to create user-centered solutions \cite{greene1987stakeholder, muller1993participatory}. We adopted this approach to develop ResonaVis with music composers, ensuring alignment with real-world compositional practices. Beyond interface design, we focused on identifying \textit{visual representations} and \textit{analytical workflows} to support interpretation of interaction data from therapeutic sessions. Following a visualization design study methodology \cite{11262783}, we iteratively refined data abstractions, tasks, and visual encodings with domain experts. Prior work in music and affective technology highlights the importance of participatory approaches for aligning systems with users’ perceptual and creative needs \cite{wilkie2013towards, grond2019participatory}.

\subsubsection{Stakeholder Collaboration}

Our team collaborated with two additional undergraduate researchers and music composers \add{as a part of the University of Delaware's (UD) Undergraduate Research Program}\new{, distinct from the Study 1 evaluation participants,} over a seven-week period through weekly design sessions. The collaboration began by eliciting composers’ existing workflows, challenges, and strategies for reasoning about musical structure. We then introduced sample interaction logs from uCue and explored how such data might inform composition. Through iterative discussions, sketching, \new{lo-fi and mid-fi prototyping,} we co-developed candidate visual representations, refined their encodings, and evaluated their usefulness for compositional reasoning tasks. This process was informed by prior work in music visualization \cite{lima2021survey}, and supported through feedback via meetings and asynchronous communication.

\subsubsection{Design Challenges in Music Composition}

The collaboration revealed recurring challenges in composing for diverse audiences, including understanding sound compatibility, balancing complexity, and interpreting listener preferences. Importantly, composers emphasized the difficulty of reasoning about \textit{temporal interaction patterns} and \textit{layer relationships} from raw data, highlighting a gap between available data and actionable insight.

Across discussions, three themes consistently emerged. First, composers struggled with \textit{relational reasoning}, particularly identifying which layers or sounds work well together and how combinations evolve over time. Second, they noted challenges in \textit{managing complexity}, where increasing the number of layers often led to diminished clarity or overwhelming auditory experiences. Third, composers expressed difficulty in \textit{understanding and validating listener preferences}, especially given the variability across individuals and contexts.

These challenges collectively point to the need for visual representations that make interaction behavior interpretable, support comparison across compositions, and provide insight into both structural and perceptual aspects of music. A comprehensive list of identified challenges is provided in
\iflabelexists{app:challenges}
  {\cref{app:challenges}}
  {the appendix of the full paper at \url{https://osf.io/xg6su}}.

% A comprehensive list of identified challenges is provided in Appendix ~\ref{app:challenges}.

\subsection{Design Goals}
Based on these challenges, we derived a set of design goals to guide the development of ResonaVis: \textit{(DG1)} the system should support reasoning about relationships between musical layers, including co-occurrence and transitions, \textit{(DG2)} it should enable analysis of temporal interaction patterns, capturing how engagement evolves over time, \textit{(DG3)} it should integrate interaction data with acoustic features to support cross-modal reasoning, \textit{(DG4)} it should facilitate comparison across sessions, songs, and participants, and \textit{(DG5)} the system should support the generation of actionable insights, enabling composers to translate observed patterns into compositional decisions.
% Building on these insights, we designed ResonaVis as a visual analytics system that supports exploration of interaction and musical data through coordinated visual representations. Our design is structured around four key principles: visual encoding of interaction patterns, cross-modal integration, analytical workflow support, and insight generation.

\subsection{ResonaVis Design}

% Guided by these goals, ResonaVis integrates multiple visual encodings and interaction techniques to support exploratory analysis of musical and behavioral data.
\new{Guided by these goals, ResonaVis integrates multiple visual encodings and interaction techniques to support exploratory analysis of musical and behavioral data (Figure~\ref{fig:teaser}).}

\subsubsection{Visual Encoding of Interaction Patterns}
\new{A central design goal was to transform raw interaction logs into visual encodings whose structure matches each analytical task, rather than encodings chosen for visual appeal alone. Guided by established task-encoding principles and refined through informal composer feedback, we selected each representation after weighing alternatives. To support reasoning about layer relationships (DG1), Sankey diagrams were chosen over chord diagrams or adjacency matrices because directed flow preserves transition order and remains legible across many co-occurring layers. Covariance matrices were favored over scatterplot matrices for representing pairwise layer co-usage, since composers prioritized rapid identification of strongly correlated pairs over precise distributional detail. To capture how engagement evolves over time (DG2), temporal session diagrams and time-series plots record layer activations over time, enabling detection of patterns such as sustained usage or rapid transitions; we aligned these plots on a shared, continuous time axis specifically to support the cross-layer and cross-song comparison required by (DG4), rather than disjoint or circular timelines that would obscure such comparisons.}
% A central design goal was to transform raw interaction logs into interpretable visual encodings that align with composers’ analytical needs. Through co-design, we developed a set of complementary representations that capture different aspects of interaction behavior. For example, Sankey diagrams encode transitions and co-occurrence between musical layers, enabling composers to identify frequently used combinations and switching behavior. Covariance matrices represent pairwise layer co-usage, supporting reasoning about harmonic and rhythmic compatibility. Temporal session diagrams and time-series plots capture layer activations and engagement over time, allowing composers to analyze temporal structure and detect patterns such as sustained usage or rapid transitions. Together, these encodings represent high-dimensional temporal interaction data in forms that map closely to composers’ mental models of musical structure and layering.

\subsubsection{Cross-Modal Visualization}
\label{sec:cross model}

% ResonaVis integrates multiple data modalities to support richer analysis, combining interaction data 
\new{To integrate interaction data with acoustic features (DG3), ResonaVis combines interaction data} (e.g., layer activations, sequences, and durations), musical data (e.g., acoustic features such as loudness, spectral centroid, and frequency distribution), and composition structure (e.g., modular layers and their variations). By coordinating these modalities within a unified visual interface, composers can relate behavioral patterns—such as engagement spikes—to underlying musical properties, including changes in timbre or spectral energy. This cross-modal integration enables analyses that would not be possible when considering a single data source in isolation.

\subsubsection{Visual Analytics Workflow}

% The system supports an exploratory visual analytics workflow that emerged from participatory design sessions. 
\new{The workflow emerged from participatory design sessions and is structured to support comparison across sessions, songs, and participants (DG4).} Composers begin by inspecting interaction data through timelines and summary visualizations to build an understanding of overall session behavior. They then detect patterns such as repeated layer usage, co-occurrence, and temporal trends, which provide insight into engagement and musical structure. This is followed by comparing behaviors across songs or participants to identify consistent preferences or variations. To deepen their analysis, composers relate interaction patterns to underlying musical features using acoustic visualizations (\new{DG3}), enabling connections between behavioral responses and properties such as timbre, pitch, and spectral characteristics. These ultimately inform compositional decisions (\new{DG5}), where composers modify or select layers based on the patterns identified \new{which reflects a shift from passive log observation to active, visualization-supported exploration.}
% Overall, this workflow reflects a shift from passive observation of raw logs to active exploration, interpretation, and decision-making supported by visualization.

\subsubsection{Supporting Insight Generation}
The combination of visual encodings and analytical workflow enables composers to generate insights that are difficult to obtain through listening or raw data alone (\new{DG5}). These insights span both interaction behavior and musical characteristics, bridging user engagement with compositional structure. For example, composers identified \textit{rhythmic stability} through layers that remained active over extended durations, as well as variations in activation frequency that reflected changes in musical dynamics. 
% Moments of increased interaction revealed \textit{engagement spikes}, often corresponding to the introduction of new layers or shifts in rhythmic intensity. Visualizations also supported recognition of frequently co-used layers, highlighting \textit{compatible layer combinations} for coherent arrangements. 
By integrating interaction data with acoustic features, composers further reasoned about \textit{pitch and timbre relationships}, using representations such as spectrograms to understand how different timbral qualities influenced engagement. 
% Similarly, analysis of \textit{spectral energy and brightness} through features such as spectral centroid and rolloff helped identify how variations in frequency distribution corresponded to sustained or reduced engagement. 
These insights - \new{reported in detail in Section \ref{sec:insights}}, directly informed compositional decisions \new{across relational, acoustic, and temporal dimensions.}
% , including selecting compatible layers, adjusting rhythmic and dynamic structure, refining timbral balance, and controlling spectral characteristics to better align with listener preferences and sensitivities.

\subsection{Implementation and Interface Design}
\label{sec:Implementation}

The dashboard incorporates ten coordinated visualizations grouped into two categories: \textit{Interaction Data} and \textit{Musical Data}. All views were implemented using Plotly\footnote{https://plotly.com/} and deployed via Dash\footnote{https://dash.plotly.com/}, enabling interactive exploration and linked analysis across representations. Design decisions were guided by established usability principles \cite{shneiderman2004designing}, ensuring consistency, responsiveness, and interpretability.

Rather than introducing novel visualization techniques, our contribution lies in the \textit{systematic integration and contextualization} of existing visual representations to support therapeutic music composition. The visualizations are designed to support three complementary analytical challenges. First, \textit{relational analysis} is supported through views such as the Sankey diagram and covariance matrix, enabling composers to reason about layer transitions and compatibility. Second, \textit{temporal analysis} is facilitated by timeline-based representations (e.g., full-session diagrams and time-series plots), which reveal interaction patterns and engagement over time. Third, \textit{acoustic analysis} is enabled through signal-based visualizations (e.g., waveplots, MFCC (\new{Mel-Frequency Cepstral Coefficients}), and spectrograms), supporting reasoning about pitch, timbre, and spectral characteristics. Together, these coordinated views enable composers to move between interaction behavior and acoustic properties, supporting both exploratory analysis and composition-oriented decision-making. A detailed description of each visualization, is provided in \iflabelexists{app:plots}
  {\cref{app:plots}}
  {the appendix of the full paper at \url{https://osf.io/xg6su}}. Color schemes (Viridis) and layout choices were refined with stakeholders to ensure accessibility and perceptual clarity, resulting in a cohesive visual analytics system.

\section{Study 1: Evaluation with Music Students}
For this preliminary study, we selected music students (undergraduate and graduate) trained in composition and music education as our target population \cite{nielsen1994usability, lazar2017research}. This group was accessible for iterative testing, possessed sufficient knowledge of music structure to interpret the visualizations, and was pedagogically oriented toward developmental contexts relevant to therapy. Their novice-to-intermediate status also made them more open to tool support, allowing us to surface usability and conceptual challenges before involving professional composers.
% in the next stage.

\subsection{Recruitment and Participants}
We recruited 8 student composers (5 females, 3 males) via mailing lists and word of mouth. They were aged between 19 and 46 years (mean = 24.75, SD = 8.83) and had formal training in music composition or music education. Six of the eight participants reported familiarity with music therapy or similar practices. The study welcomed participants of all genders, ethnicities, and backgrounds. Importantly, there were no specific medical conditions or exclusion criteria beyond the stated requirements. The study design was approved by the Institutional Review Board (IRB) of \add{UD} (Protocol \add{\#2097325}), ensuring that all research activities were conducted in accordance with ethical standards. Informed consent was obtained from all participants, and individual identities were not linked to the data used in the analysis. All participant data was anonymized. Table \ref{tab:participants} gives an overview of the participants. The session with P2 was a virtual Zoom session, conducted based on their availability. We also removed data for one participant (P5) from our analysis, as the session was canceled midway due to technical difficulties. Thus, our results focus on the seven participants who successfully completed the study session.

\begin{table}[t]
\centering
\footnotesize
\renewcommand{\arraystretch}{1.0}
\setlength{\tabcolsep}{2.5pt}
\rowcolors{2}{gray!10}{white}
\caption{\textbf{Participant demographics and background.}}
\label{tab:participants}

\begin{tabularx}{\columnwidth}{
>{\centering\arraybackslash}p{0.45cm}
>{\centering\arraybackslash}p{0.45cm}
>{\centering\arraybackslash}p{0.55cm}
>{\centering\arraybackslash}p{0.65cm}
>{\raggedright\arraybackslash}X
>{\centering\arraybackslash}p{0.7cm}
>{\centering\arraybackslash}p{0.6cm}}
\hline
\textbf{ID} & \textbf{G} & \textbf{Age} & \textbf{Deg} & \textbf{Major} & \textbf{Music} & \textbf{Vis} \\
\hline

P1 & F & 46 & BSc & English (Music minor) & Both & N \\
P2 & F & 22 & MSc & Linguistics & Edu & Y \\
P3 & F & 19 & BSc & Music Perf./Comp. & Both & Y \\
P4 & F & 20 & BSc & Music + Psychology & Comp & Y \\
P5 & F & 20 & BSc & Cognitive Sci., Viola Perf. & Both & N \\
P6 & M & 22 & MSc & Music Composition & Both & Y \\
P7 & M & 24 & MSc & Music Performance & Both & N \\
P8 & M & 25 & MSc & Music Performance & Both & N \\

\hline
\end{tabularx}
\end{table}

\subsection{Study Design}
\label{sec:stude_design}
\new{We follow a design study approach, grounding system development in domain-specific needs and iterative collaboration with composers, consistent with established visualization design study methodologies \cite{sedlmair2012design}. We combine qualitative and quantitative metrics, focusing on how well composers could interpret the visualizations and their overall satisfaction with the interface. To keep sessions under one hour, we selected a subset of visualizations for the structured evaluation tasks, focusing on relational and acoustic analysis (Section~\ref{sec:Implementation})—covering rhythmic attack, pitch, frequency, amplitude, overtone, brightness, darkness, timbre, and instrument type—based on discussions with our research team and stakeholders. Temporal analysis visualizations (e.g., full-session diagrams, time-series plots) remained available but were not the focus of a dedicated evaluation task; their value was instead assessed qualitatively through usability feedback (Section~\ref{sec:usability}). Participants received verbal definitions and examples for shared understanding (sheet in Supplementary Materials). We did not test clinical outcomes, but explored whether the dashboard could scaffold composition decisions aligned with therapeutic goals. Task 1 focused on Interaction Data, and responses from this task served as input for Task 2.}
% We follow a design study approach, grounding system development in domain-specific needs and iterative collaboration with composers, consistent with established visualization design study methodologies \cite{sedlmair2012design}. We combine qualitative and quantitative metrics, focusing on how well composers could interpret the visualizations and their overall satisfaction with the interface. To keep sessions under one hour, we selected a subset of visualizations, covering rhythmic attack, pitch, frequency, amplitude, overtone, brightness, darkness, timbre, and instrument type, based on discussions with our research team and stakeholders. Participants were verbally given definitions and examples to ensure shared understanding (a shared terminology sheet is included in our Supplementary Materials). The study did not test clinical outcomes, but explored whether the dashboard could scaffold composition decisions aligned with therapeutic goals. Task 1 focused on Interaction Data, and responses from this task served as input for Task 2.

During the study, participants interacted with two sections of the interface: the Interaction and Audio section, which included the Participant Analysis (S1) and Comparative Analysis (S2) interfaces, as shown in Figure \ref{fig:Study}. In S1, participants were required to select a song of their choice, choose the number of participants to include (with a preference for selecting all to obtain a net average), pick a plot for the challenge, and select the appropriate options from a dropdown menu. The observations made in S1 were then used as input for S2. In S2, participants selected the same song chosen in S1, selected a plot for the challenge, and input their observations from S1 into two comparative plots provided in this section. Finally, they were asked to write down their findings in response to specific questions related to their observations (see Supplementary Materials for the entire study design). Table \ref{tab:evaluation_questions} describes the three design challenges which we formulated collaboratively with our stakeholders using language commonly employed in music therapy  \cite{tsiris2020impact}. These design challenges were used to assess the participants' understanding and the dashboard's ability to facilitate analysis. We calculate the raw NASA-TLX workload scores \cite{hart2006nasa} after each challenge which captured participants’ perceived cognitive and emotional workload, and the System Usability Score (SUS) \cite{vlachogianni2022perceived} at the end to determine the overall usability.

\begin{figure}[t]
\centering
\includegraphics[width=1\columnwidth]{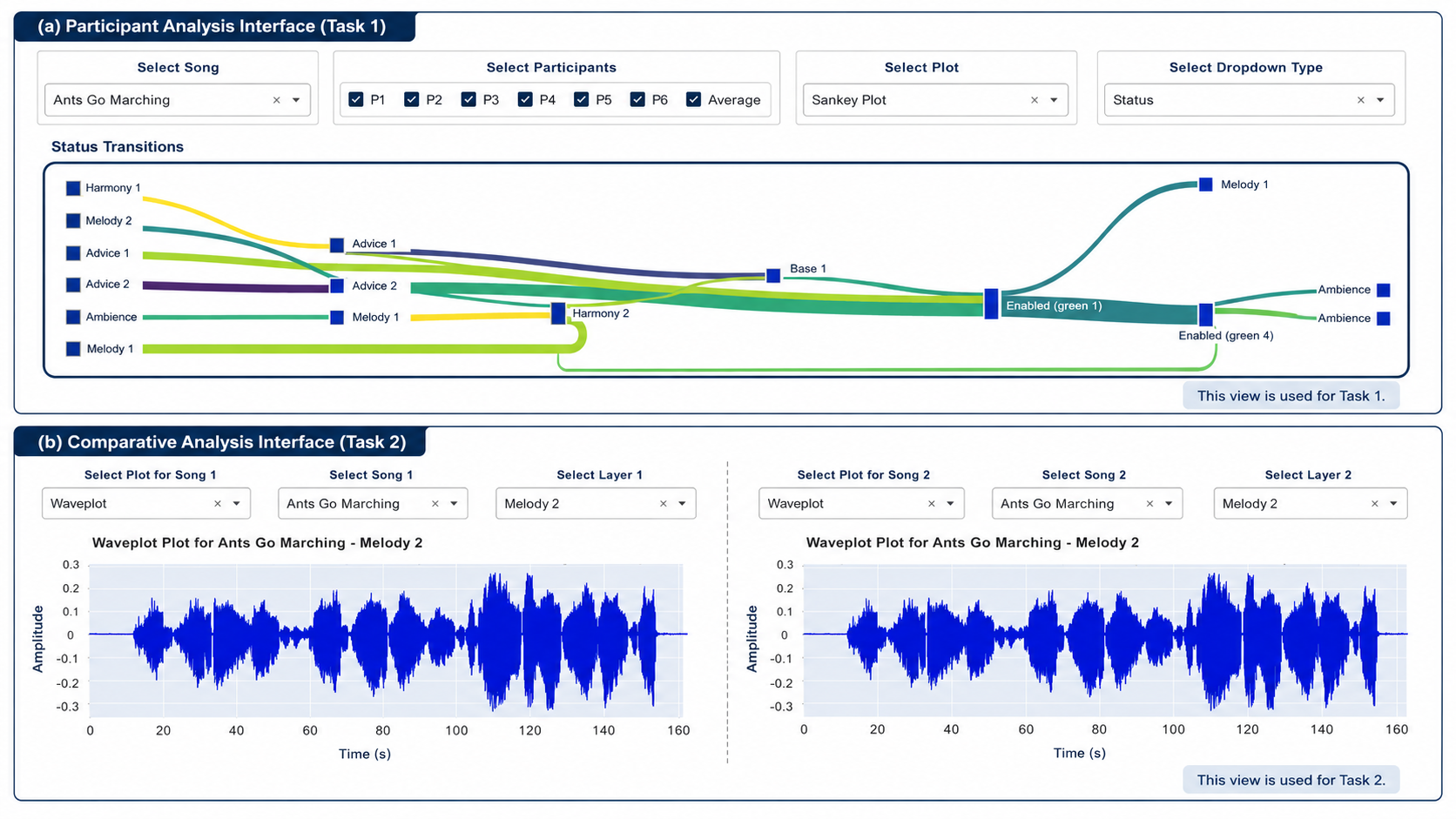}
% \caption{\textbf{Study interface.} 
% Section S1 (Participant Analysis) enables exploration of individual interaction patterns using visualizations (e.g., Sankey diagrams). Section S2 (Comparative Analysis) supports side-by-side comparison across layers, songs, or participants (e.g., waveplots).}
% \caption{\textbf{Study interface.} 
\caption{\textbf{Study interface used in evaluation tasks.}
\new{(a) Participant Analysis Interface (Task 1): users select a song, participants, plot type, and category to explore individual interaction patterns; shown here, a Sankey diagram of status transitions.}
\new{(b) Comparative Analysis Interface (Task 2): users independently select a plot, song, and layer per panel for side-by-side comparison; shown here, waveplots for the same song and layer on a shared temporal axis (0--160s)}.}
\label{fig:Study}
\end{figure}

% \begin{table}[t]
% \centering
% \rowcolors{2}{gray!10}{white}
% \renewcommand{\arraystretch}{1.15}
% \caption{\textbf{Evaluation design challenges and associated visualizations.}}
% \label{tab:evaluation_design challenges}

% \begin{tabularx}{\columnwidth}{>{\centering\arraybackslash}p{0.6cm} X >{\raggedright\arraybackslash}p{2.4cm}}
% \hline
% \textbf{Q} & \textbf{Evaluation Question} & \textbf{Visualizations} \\
% \hline

% \textbf{1} & How do the rhythmic and dynamic characteristics of selected layers complement each other? & Covariance Matrix (S1), Waveplot (S2) \\

% \textbf{2} & How does the popularity of frequently used layers relate to spectral energy distribution and perceived sound brightness? & Bar Graph (S1), Spectral Rolloff (S2), Spectral Centroid (S2) \\

% \textbf{3} & How do transitions between layers correspond to changes in frequency and pitch over time, and how might they influence the listening experience? & Sankey Diagram (S1), Mel Spectrogram (S2) \\

% \hline
% \end{tabularx}

% \end{table}
\begin{table}[t]
\centering
\rowcolors{2}{gray!10}{white}
\renewcommand{\arraystretch}{1.0}
\small
\caption{\textbf{Design Challenges and associated visualizations.}}
\label{tab:evaluation_questions}
\begin{tabularx}{\columnwidth}{>{\centering\arraybackslash}p{0.4cm} X >{\raggedright\arraybackslash}p{2.1cm}}
\hline
\textbf{Q} & \textbf{Evaluation Question} & \textbf{Visualizations} \\
\hline
\textbf{1} & How do rhythmic/dynamic characteristics of selected layers complement each other? & Covariance Matrix (S1), Waveplot (S2) \\
\textbf{2} & How does popularity of frequently used layers relate to spectral energy and perceived brightness? & Bar Graph (S1), Spectral Rolloff/Centroid (S2) \\
\textbf{3} & How do layer transitions correspond to frequency/pitch changes over time, and how might they affect listening experience? & Sankey Diagram (S1), Mel Spectrogram (S2) \\
\hline
\end{tabularx}
\end{table}

\subsection{Procedure}

Participants engaged with ResonaVis through multiple coordinated visualizations. Each session lasted approximately one hour, and participants took part only once. Sessions were conducted in person at the Sensify Lab in UD for all participants except P2, who participated remotely. In-person sessions were conducted on a 24-inch monitor (1920×1080 resolution), while the remote participant was instructed to use the largest available screen and share their display. All sessions were audio recorded with participants’ consent.

The session began with a brief introduction to the dashboard, during which participants were guided through its features and interaction mechanisms. They were encouraged to freely explore the interface to become familiar with the visualizations before proceeding. This was followed by a short pre-session interview, where participants described their typical music composition process, including how they develop and refine musical ideas. Participants then completed a pre-session questionnaire assessing their confidence in composing music for children with ASD and their ability to interpret visual data. Following this, they engaged in the three design challenges designed to evaluate the dashboard. These design challenges required participants to analyze visualizations, extract insights, and relate these insights to potential compositional decisions. After each design challenge, participants documented their observations and provided feedback on the usefulness and clarity of the visualizations.To assess cognitive workload, participants completed the NASA-TLX after each challenge. At the conclusion of all challenges, they completed the SUS test to evaluate overall usability. Participants then repeated the confidence questionnaire to capture changes in perceived ability following interaction with the system. Finally, sessions concluded with a short debriefing, where participants reflected on their experience, discussed challenges encountered, and suggested potential improvements.

\subsection{Data Collection and Analysis}
In this study, data analysis was conducted using both quantitative and qualitative approaches. Quantitative analysis involved statistical examination of metrics such as  SUS scores, and the Raw Task Load Index (scale of 10\new{, following Hart~\cite{hart2006nasa}, which validates its reliability}) for each task. These metrics provided measurable insights into participant performance and the overall usability of the dashboard. On the qualitative side, thematic analysis \cite{clarke2017thematic} was employed to analyze feedback from participants and their responses to study questions. This analysis helped identify common themes and suggestions, offering deeper insights into the participants' experiences and guiding potential improvements to the dashboard. The responses of the participants were double coded by the first author along with the music composers on the team. 
% specializing in music composition for therapy. 
\subsection{Results}

The results are organized into subsections covering pre-session interviews, task performance, insights from visualizations, workload and usability scores, confidence changes, and qualitative feedback. Together, these findings provide a holistic view of how participants engaged with the visualisations and ResonaVis. A post-session interview further examined whether their perspectives on composing for children with ASD had shifted.

\subsubsection{Pre-Session Interview Results}

We conducted a pre-session interview to understand participants’ existing approaches to music composition and the decision-making processes underlying their work. Participants were asked to describe their typical workflow, from initial inspiration to refinement, including how they iteratively develop and evaluate their compositions. 
Analysis of the responses revealed three primary compositional approaches. First, some participants adopted an \textit{inspiration-driven approach}, drawing from familiar musical pieces and using emulation or adaptation as a starting point (P1). Second, a majority of participants described a \textit{segmented and iterative workflow}, in which individual musical elements (e.g., rhythm, melody, or harmony) were composed separately and refined through repeated listening, visual cues, and feedback (P2, P3, P6, P7, P8). Third, one participant emphasized a \textit{therapeutic alignment approach}, revisiting and adjusting compositions to better match emotional goals and principles from music therapy (P4)---\new{such as reducing sensory overload and promoting calmness through predictable rhythmic and tonal structures}. Across these approaches, participants relied heavily on subjective judgment, prior experience, and iterative listening, with limited use of structured data or visual analysis to guide their decisions. This suggests that, prior to using ResonaVis, compositional strategies were largely experience-driven and heuristic-based rather than explicitly informed by multi-dimensional analysis of musical features. These baseline insights provide an important point of comparison for subsequent results, enabling us to examine how participants’ workflows evolved from intuition-driven practices to more data-informed and adaptive compositional strategies after interacting with the system.

\subsubsection{Insights Derived}
\label{sec:insights}
\new{We conducted a thematic analysis of participants' written responses, \new{organized around the three analytical dimensions from Section~\ref{sec:Implementation}: relational, acoustic, and temporal analysis. As noted in Section~\ref{sec:stude_design}, structured evaluation tasks addressed relational and acoustic dimensions; temporal analysis is discussed qualitatively below.}
First, \textit{\new{relational} reasoning} emerged through \new{compatible layer combinations and transitions.} \new{Covariance-based visualizations enabled participants to interpret co-occurrence patterns as indicators of rhythmic alignment and dynamic complementarity, while Sankey diagrams supported interpretation of flow thickness and structure to understand how layer changes contributed to variation, contrast, and perceived smoothness in the auditory experience.}
Second, \textit{\new{acoustic} reasoning} was observed as participants connected frequency-based representations (e.g., spectral centroid) to perceptual qualities such as calmness, intensity, and brightness\new{, and related spectral energy distributions to sustained or reduced engagement.}
\new{Third, while \textit{temporal reasoning} was not the focus of a dedicated evaluation task, participants qualitatively reported that timeline-based views helped them identify engagement patterns over time that would be difficult to perceive through listening alone.}
Importantly, participants did not treat these dimensions in isolation. Instead, they integrated \new{acoustic, relational,} and dynamic cues to form holistic interpretations of musical structure and perception. This suggests that ResonaVis supports \textit{multi-dimensional reasoning}, enabling users to connect low-level acoustic features with higher-level perceptual and therapeutic interpretations.}

\new{\textbf{Role of ResonaVis: } These insights were closely tied to the visual encodings provided by the system. \new{Covariance and Sankey diagrams supported relational reasoning about co-occurrence and transitions, while spectral plots enabled acoustic interpretation of energy distribution and brightness.} Together, these coordinated views supported users in constructing meaningful interpretations of complex musical relationships. Qualitatively, rhythmic alignment and spectral energy were among the most frequently observed themes, while transition-related insights captured more holistic aspects of the auditory experience. These findings align with observed confidence increases (Section~\ref{sec:conf}) and NASA-TLX (Section~\ref{sec:nasa}) workload patterns, suggesting that the cognitive effort contributed to deeper understanding rather than overload. A detailed codebook is provided in \iflabelexists{app:codebook}
  {\cref{app:codebook}}
  {the appendix of the full paper at \url{https://osf.io/xg6su}}.}

\subsubsection{Usability Results}
\label{sec:usability}
\new{We evaluated the usability of ResonaVis using the System Usability Scale (SUS). The interface achieved a mean SUS score of 70.7 (SD = 17.8) in its first iteration, exceeding both the average score for initial systems (62) and the overall benchmark for web-based interfaces (68.05) reported by Bangor \textit{et al.}~\cite{bangor2008empirical}, placing it in the \textit{Good} range on associated adjective rating scale, above \textit{OK} and approaching \textit{Excellent}. This indicates that participants generally perceived the system as usable and approachable, even at an early stage of development. However, scores varied substantially across participants, ranging from 42.5 to 92.5. While several participants (e.g., P6: 92.5, P8: 85.0) reported high usability and found the system intuitive and well-integrated, others (e.g., P2: 42.5, P7: 52.5) indicated challenges in using the interface. \new{Across analytical dimensions, usability varied by visualization type: relational views, particularly the Sankey diagram, required additional explanation before participants could confidently interpret layer transitions, whereas bar graphs and the covariance matrix were comparatively intuitive to read. Acoustic visualizations were generally accessible, supporting spectral reasoning with minimal guidance. Temporal views, while not formally evaluated in structured tasks, received qualitative endorsement: participants reported that full-session diagrams and time-series plots surfaced engagement patterns difficult to perceive through listening alone.} This variability suggests that while the system is effective for some users, it may require additional support or refinement to ensure consistent usability, particularly for those less familiar with interpreting visualizations or multi-dimensional musical data. These observations align with the improvements in confidence (Section~\ref{sec:conf}) and the moderate cognitive demands observed in the NASA-TLX results (Section~\ref{sec:nasa}), suggesting that the system supports meaningful analytical engagement even when usability is not uniformly optimal. Overall, the SUS results indicate that ResonaVis provides a strong foundation for usability in its current form, while also highlighting opportunities for improving learnability, consistency, and guidance to better support a wider range of users.}

\subsubsection{NASA-TLX for Each Challenge}
\label{sec:nasa}

We used NASA-TLX to assess perceived workload across tasks and identify opportunities to improve ResonaVis.

\textbf{Challenge 1: }
Participants reported high mental demand (5.86) and effort (5.43), with low physical (0.43) and temporal demands (2.43). Despite this, performance was high (7.86) and frustration remained low (2.71), suggesting that while the task required sustained reasoning, participants could engage effectively without negative effect. This indicates that ResonaVis supports cognitively intensive compositional tasks while maintaining usability.

\textbf{Challenge 2: }
Mental demand (5.43) remained similar to Challenge 1, but effort (5.00) and frustration (3.00) increased, alongside a drop in performance (6.43). Physical (0.43) and temporal demands (1.57) stayed low. This suggests greater cognitive complexity or ambiguity in interpreting the visualizations. The increased effort and frustration point to opportunities for improved guidance for interaction support.

\textbf{Challenge 3: }
All workload dimensions were low, including mental (2.57), physical (0.29), and temporal demands (1.00), while performance remained high (7.57). Effort (3.29) and frustration (2.57) were also minimal, indicating that the task required little cognitive overhead. This reflects strong baseline usability for simpler tasks.

\textbf{Cross-Challenge Insights: }
A clear gradient emerges: Challenges 1 and 2 required higher cognitive engagement, while Challenge 3 was lightweight. Notably, high mental demand in Challenge 1 did not reduce performance or increase frustration, indicating productive cognitive load. In contrast, Challenge 2 suggests that some forms of complexity are less well supported, highlighting the need for better interpretability and scaffolding. Figure \ref{fig:nasa_tlx} summarizes these results.

\textbf{Relation to Confidence Gains: }
These workload patterns contextualize the significant confidence gains (Section \ref{sec:conf}). Although Challenges 1 and 2 required moderate to high cognitive effort, participants still showed increased confidence across all dimensions. This suggests that the effort was productive, supporting learning and deeper engagement. Rather than minimizing cognitive load, effective visual analytics systems should help users manage and leverage it.

\begin{figure}[t]
    \centering
    \includegraphics[width=\columnwidth]{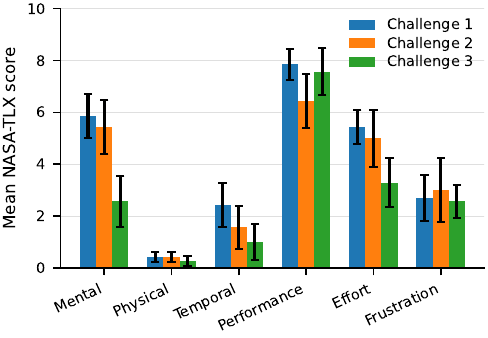}
    \caption{\textbf{NASA-TLX scores across the three challenges.} Error bars indicate standard error. }
    % Challenges 1 and 2 exhibit relatively higher mental demand and effort, while Challenge 3 shows lower perceived workload on these dimensions. }
    \label{fig:nasa_tlx}
\end{figure}

% \begin{figure}[h]
%     \centering
%     \begin{subfigure}[b]{0.33\textwidth}
%         \includegraphics[width=\textwidth]{Figures/Load Test/TLI1.png}
%         \caption{\textbf{Challenge 1}}
%         \label{fig:challenge1}
%     \end{subfigure}
%     \hfill
%     \begin{subfigure}[b]{0.33\textwidth}
%         \includegraphics[width=\textwidth]{Figures/Load Test/TLI2.png}
%         \caption{\textbf{Challenge 2}}
%         \label{fig:challenge2}
%     \end{subfigure}
%     \hfill
%     \begin{subfigure}[b]{0.33\textwidth}
%         \includegraphics[width=\textwidth]{Figures/Load Test/TLI3.png}
%         \caption{\textbf{Challenge 3}}
%         \label{fig:challenge3}
%     \end{subfigure}
%     \caption{\textbf{Task Load Index for each Challenge}}
%     \label{fig:task_load_index}
% \end{figure}
% \FloatBarrier

% Overall, the usability and task performance results indicate that the tool was generally well-received, with a moderate to high usability score and manageable task loads across challenges. While some participants experienced difficulties, most found the tool easy to use with varying levels of cognitive and emotional demands depending on the challenge. 

% \newpage

\subsubsection{Qualitative Feedback}
\label{sec:qual}

We analyzed participants’ reflections on their compositional strategies before and after using ResonaVis to understand how the system influenced their approach to designing music for individuals with ASD. The analysis reveals a clear shift from \textit{rule-based} to \textit{adaptive} composition strategies. Prior to using the system, participants largely relied on conservative heuristics grounded in assumptions about sensory sensitivities. Across themes, compositions emphasized simplicity, predictability, and reduced sensory complexity, including steady rhythms, limited dynamic variation, mid-to-low pitch ranges, and less bright timbres. These strategies reflected a precautionary approach aimed at avoiding overstimulation. After interacting with ResonaVis, participants demonstrated a transition toward \textit{data-informed exploration}. Rather than adhering to static rules, they began incorporating controlled variation in rhythm, dynamics, timbre, and spectral properties based on observed interaction patterns and visual insights. Participants experimented with rhythmic transitions, and actively balanced spectral energy to achieve desired perceptual effects. Importantly, this shift did not reflect a move toward complexity alone, but toward \textit{informed flexibility}. Participants continued to consider sensory sensitivity but adapted their decisions dynamically based on context. This transition from rigid heuristics to nuanced, context-aware reasoning suggests that ResonaVis supports more reflective and responsive compositional practices.

\textbf{Cross-Participant Trends: }
Across participants, a consistent pattern emerged: initial strategies were precautionary and generalized, whereas post-session approaches were more individualized and adaptive. Participants reported increased willingness to experiment, refine, and iterate on compositional elements, indicating a shift toward more context-aware music design. These qualitative changes align with the observed increases in confidence (Section~\ref{sec:conf}) and NASA-TLX workload patterns, suggesting that the cognitive effort required by the system contributed to deeper understanding rather than overload. A detailed breakdown of these changes is provided in \iflabelexists{app:qual_feedback}
  {\cref{app:qual_feedback}}
  {the appendix of the full paper at \url{https://osf.io/xg6su}}.

% A detailed breakdown of these changes across rhythmic, timbral, and spectral dimensions is provided in Appendix~\ref{app:qual_feedback}.

% % \textbf{Relation to Quantitative Findings: }
% These qualitative changes align with the observed increases in confidence (Section~\ref{sec:conf}) and NASA-TLX workload patterns, suggesting that the cognitive effort required by the system contributed to deeper understanding rather than overload. A detailed breakdown of these changes across rhythmic, timbral, and spectral dimensions is provided in Appendix~\ref{app:qual_feedback}.
% % Together, these findings reinforce the role of ResonaVis as a tool that supports both analytical reasoning and informed creative decision-making. 
% % \FloatBarrier

% \newpage
\subsubsection{Changes in Confidence Level}
\label{sec:conf}

Participants’ confidence in composing music for individuals with ASD (0–5 scale) increased significantly after using ResonaVis. Confidence was evaluated across four dimensions: (i) rhythmic and dynamic elements, (ii) pitch and timbre, (iii) spectral energy and brightness, and (iv) interpretation of graphs and data.

% Across all dimensions, participants reported statistically significant improvements. 
Mean confidence increased across all four dimensions. Confidence in working with rhythmic and dynamic elements increased from 2.86 (SE = 0.34) to 4.14 (SE = 0.26; $p = 0.0004$), representing the largest gain. Similarly, confidence in pitch and timbre improved from 2.86 (SE = 0.51) to 3.86 (SE = 0.34; $p = 0.038$), while confidence in reasoning about spectral energy and brightness increased from 2.29 (SE = 0.42) to 3.57 (SE = 0.38; $p = 0.004$).  Participants also became more confident in interpreting visual representations of musical data, with scores increasing from 3.07 (SE = 0.22) to 4.14 (SE = 0.26; $p = 0.032$). This improvement suggests that ResonaVis not only supported compositional reasoning but also helped users develop the ability to read and make sense of complex visual encodings.

While the sample size is small and confidence was measured using a Likert scale, \add{the consistent direction of the mean changes indicates a positive trend}. \new{We note that participants also received verbal terminology training prior to the tasks (Section~\ref{sec:stude_design}), and we cannot fully disentangle gains attributable to this exposure from gains attributable to interacting with the visualizations themselves; we view confidence as reflecting the combined effect of terminology familiarity and visual analytical practice}. These results should therefore be interpreted as indicative rather than conclusive, motivating further validation with a larger and more diverse participant pool. Overall, the gains across all four dimensions indicate that ResonaVis facilitated a broader understanding of acoustic features and their role in therapeutic music composition. Notably, improvements in both compositional aspects (e.g., rhythm, timbre, spectral features) and analytical skills (e.g., graph interpretation) suggest that the system supports both creative and analytical engagement. Figure~\ref{fig:Confidence} illustrates these changes in confidence before and after using ResonaVis.

% \begin{figure}
% \centering
% \includegraphics[width=\columnwidth]{figs/Confidence/confidence_pre_post_vgtc_plot_bottomright.pdf}
% \caption{\textbf{Participants' confidence before and after using ResonaVis.} Points indicate mean confidence scores across four composition-related dimensions, and error bars indicate standard error. Asterisks denote statistically significant pre- to post-session improvements ($p < 0.05$).}
% \label{fig:Confidence}
% \end{figure}
% % \FloatBarrier

\begin{figure}
\centering
\includegraphics[width=\columnwidth]{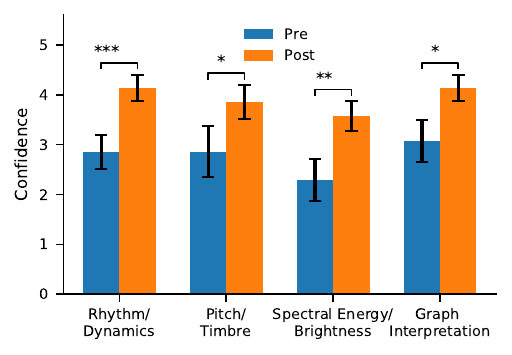}
\caption{\textbf{Pre–post confidence gains:} Error bars show standard error; * indicate statistical significance (* p < 0.05, ** p < 0.01, *** p < 0.001).}
\label{fig:Confidence}
\end{figure}

% (* p < 0.05, ** p < 0.01, *** p < 0.001)

\subsubsection{Post-Session Interview Results}

Following the study, participants reflected on how the visualizations aligned with or differed from their existing compositional practices, and whether these representations influenced their approaches to developing and refining music. Overall, participants reported that the visualizations provided new and actionable perspectives on composition. Several participants described how the system supported more informed decision-making prior to composing. For example, P3 noted that the visualizations enabled them to pre-select instruments and identify compatible layers for mixing, particularly when aiming to create soothing auditory experiences. Similarly, P6 and P8 highlighted the usefulness of comparative analysis features, which allowed them to quickly identify differences between layer combinations. P6 further indicated a strong intention to incorporate the tool into their regular compositional workflow, suggesting that ResonaVis can integrate effectively into existing creative practices for some users. At the same time, participants’ responses revealed variability in how well the system aligned with individual compositional styles. While P1, P2, and P7 acknowledged the potential value of the tool, they indicated that they might not frequently use it due to differences between their established workflows and the visualization-driven approach. For instance, P7 emphasized a preference for composing through musical notation, suggesting that the absence of note-level representations limited the tool’s relevance for their practice. However, participants also noted that increased familiarity with the visualizations, or alignment between the tool and their compositional approach, could improve its utility.

\textbf{Cross-Participant Insights: }
Across responses, a key pattern emerges: ResonaVis is particularly effective as a \textit{pre-compositional and analytical support tool}, helping users explore layer relationships, evaluate combinations, and reason about perceptual outcomes before or during composition. Rather than replacing existing workflows, the system appears to augment them by providing additional perspectives that are difficult to obtain through listening alone.

\textbf{Implications for Adoption: }
These findings suggest that the usefulness of ResonaVis depends on the degree of alignment between its visualization-driven paradigm and the user’s existing compositional style. For participants open to integrating visual and data-driven reasoning, the system supported deeper insight and workflow integration. For others, particularly those relying on notation-based or highly internalized processes, the system was perceived as complementary but not essential. This highlights the importance of designing flexible interfaces that can accommodate diverse compositional practices. 
% These results indicate that ResonaVis not only supports analytical understanding of musical structure but also has the potential to influence compositional decision-making, particularly in early-stage exploration and planning.

\section{Study 2: Observational Case Study}
This case study explored ResonaVis within the workflows of two experienced composers (P9 and P10) who have created music for children with ASD and were recruited via word of mouth and through the research team's existing contacts (\add{IRB Protocol \add{\#2097325})}. By observing their interactions with the system and gathering reflections, we examined how engagement-driven visualizations could inform advanced compositional decisions and validate their design choices.
% from earlier co-design work.

\subsection{Method}
% \new{We invited a music composer who had previously composed and produced songs in our modular musical library for a virtual 1-hour session. The participant was female, age 21, and contributed to the creation of uCue’s musical library. }
% We invited a 21-year-old female composer who had previously contributed to the creation of a modular musical library developed by prior work done by Karwankar \textit{et al.} \cite{10.1145/3713043.3727053} for a virtual 1-hour session. During the observation session, we introduced the ResonaVis’s features and visualizations, followed by a task where the composer explored interaction logs and musical features from past sessions. The participant was encouraged to verbalize her thought process, identify patterns, and propose potential compositional adjustments. We recorded both screen activity and audio commentary. These recordings were transcribed and thematically analyzed to capture recurring insights, usability issues, and reflections on compositional strategies.
We conducted two one-hour sessions with experienced composers.  P9 (21, Female) was a composer who had previously contributed to the creation of a  musical library for children with ASD. P10 (Age not reported, Female) was an experienced music composer and director who has composed for children with ASD. She was pursuing a Master’s degree in piano performance and had no prior experience with data visualization tools. P9 participated in a virtual session, while P10 was an in-person session. Participants were introduced to ResonaVis’s features and visualizations, followed by a task where they explored interaction logs and musical features (Study design in Supplementary Material). They were encouraged to verbalize their thought processes, identify patterns, and propose compositional adjustments. Both sessions were audio recorded and transcribed for thematic analysis.

\subsection{Findings}
% Our observations and interview insights are organized into three themes:

% \subsubsection{Baseline Confidence and Composition Approach.}
% In the pre-interview, the participant reflected that when she first began composing for children with ASD, her confidence was very low due to a lack of knowledge about their musical preferences. Her early approach emphasized keeping melodies simple, making each layer distinct, and relying on intuition. She noted that, at the time, she did not consider many potentially relevant musical factors, such as how variations in tempo or transitions might help guide attention or activity, because she was unaware of them. Her sole aim was to make the song as calming as possible.
\subsubsection{Baseline Confidence and Composition Approach.}
Both participants reflected on the challenges of composing for children with ASD, particularly in the early stages of their practice. P9 described initially relying on intuition and focusing on simple, calming compositions without explicitly considering how different musical features might influence engagement over time. Similarly, P10 reported low initial confidence when beginning to compose for this audience, noting uncertainty about what musical elements were appropriate. To address this, she relied on prior research on sensory preferences, particularly differences in high, medium, and low frequency ranges. Her compositions emphasized simplicity and clarity of layers, with a primary goal of producing calming musical experiences. However, like P9, she noted that aspects such as transitions and temporal dynamics were not systematically considered in her early work.

\subsubsection{Experience Using ResonaVis.}  
\label{sec:exp}
After being introduced to ResonaVis, both participants found the system useful in supporting compositional reasoning, though in slightly different ways based on their backgrounds.
P9’s responses aligned closely with findings from Study~1, particularly in using engagement patterns to reason about when to introduce or delay layers to shape musical structure. She highlighted how the system consolidated “a lot of information” in one place, enabling efficient review of interaction logs without replaying full recordings and supporting rapid “what-if” exploration.
P10, despite having no prior experience with data visualization, responded positively to the system and found it valuable for practical composition. She particularly appreciated the bar charts and covariance matrix, noting that they clearly conveyed which layers were most used and how different musical elements co-occurred. She was also able to relate the frequency distributions shown in the mel spectrogram to her compositional decisions, using them to reason about pitch ranges and refine transitions between sections. Reflecting on this process, she noted, \textit{“I can almost imitate the sounds while looking at the transitions in the Sankey Chart (While imitating some musical transitions after this quote)”} indicating how the visualizations supported a mapping between visual patterns and auditory expectations.
Across both participants, integrating interaction logs and audio features enabled efficient identification of patterns such as frequently activated layers and transitions, supporting more deliberate decisions about when to introduce, delay, or combine layers to shape the composition’s energy and flow.
% Across both participants, the integration of interaction logs and audio features enabled efficient identification of patterns such as frequently activated layers and transition points. This supported more deliberate decisions about when to introduce, delay, or combine layers to shape the overall energy and flow of a composition.

% \subsubsection{Perceived Value, Usability, and Suggestions.}
% The participant rated the system’s usability at 77.5 (SUS), indicating good usability. 
% % She stated, \textit{“If I had such a tool with me four years back when I started composing for this audience, I would use it frequently. It would help me cater the music for children with autism more confidently and independently, taking a leadership role in music composition.”} 
% She reflected, \textit{``When I first started composing for this audience, I often lacked confidence in knowing whether my choices were appropriate. A tool like this could have provided useful guidance and helped me make decisions more independently.''} She suggested adding accessible definitions or hover-based explanations for each visualization, noting that many musicians are unfamiliar with data visualization techniques, and such support could improve adoption. Also she emphasized the idea about a collaboration platform which would help musicians to start a discussion to facilitate collaborate decision making in a shared space.
\subsubsection{Perceived Value, Usability, and Suggestions.}
\label{sec:perc}
Both participants perceived ResonaVis as a useful tool for supporting therapeutic composition. P9 emphasized that such a system could have improved her confidence and independence when she first began composing for this audience, stating, \textit{“When I first started composing for this audience, I often lacked confidence in knowing whether my choices were appropriate. A tool like this could have provided useful guidance and helped me make decisions more independently.”}

In terms of usability, P9 rated the system at 77.5 (SUS), while P10 rated it at 70, resulting in an average usability score of 73.75, indicating good overall usability. P10’s rating is notable given her lack of prior experience with visualization tools. She described the system as practical and applicable to real-world composition workflows, noting \textit{“It feels like a practical tool I could actually use in my composition process, not something abstract.”} At the same time, P10 reported some difficulty in interpreting certain visualizations (\new{similar to section \ref{sec:usability}}), highlighting a learning curve for users without technical backgrounds. This aligns with P9’s suggestion to include clearer explanations or onboarding support. Both participants emphasized the importance of making visualizations more accessible to musicians who may not be familiar with data-driven tools.

\subsection{Summary}
Overall, the case study demonstrates that ResonaVis can support both experienced and non-technical composers by bridging domain knowledge with data-driven insights. While P9 leveraged the system to refine existing strategies, P10 used it to ground her compositional decisions in visual evidence, particularly in relation to frequency and transitions. Together, their feedback highlights the system’s practical value, while also underscoring the need for improved interpretability to support broader adoption among musicians.

% \subsection{Summary}
% Overall, the case study showed that the ResonaVis helped the composer efficiently review session data, avoid replaying full recordings, and explore “what-if” scenarios when considering musical adjustments. Her reflections indicated that the visualizations consolidated otherwise scattered information, supporting more deliberate decisions about engagement, transitions, and instrumentation. While she suggested improvements such as clearer definitions for visualizations, her feedback highlighted the tool’s potential to scaffold confidence and independence in therapeutic composition.

% \subsection{Challenges}
% \new{The primary challenge in the observational study was that we could not evaluate the complete music composition process, as producing a single track often spans weeks or even months. Our observations, therefore, focused on early-stage ideation and engagement analysis using MusicVis, which may not fully reflect the tool’s long-term impact on final compositions. Additionally, because the composer was already familiar with the song and therapeutic context, her interpretations may differ from those of composers working with unfamiliar material, limiting generalizability.}

\section{Discussion}

We explore how visualization supports data-driven music composition by enabling composers to interpret patterns and generate insights. Rather than evaluating clinical outcomes, we focus on how visual representations shape compositional reasoning and decisions. 
% We revisit our research questions by synthesizing findings across studies and prior work.
% ResonaVis investigates how visualization can support data-driven music composition decisions for children with ASD by enabling composers to interpret patterns in interaction and acoustic data. Rather than evaluating clinical outcomes, our goal was to understand how visual representations shape compositional reasoning and decision-making. We revisit our research questions by synthesizing findings across studies and situating them within existing literature.

\subsection{RQ1: Visualization-Supported Compositional Insights}

Our findings show that visualization enables composers to reason about musical structure through observable patterns in rhythm, spectral properties, and temporal transitions (Sections \ref{sec:insights} and \ref{sec:qual}). Participants consistently identified relationships such as rhythmic alignment across layers, balance in spectral energy and brightness, and structured transitions between musical elements. These insights were not derived from listening alone, but from recognizing patterns in visual encodings such as timelines, covariance matrices, and flow-based representations.

These observations align with prior work highlighting the importance of rhythmic predictability, timbral balance, and low-brightness sound textures in therapeutic music contexts \cite{allen2009hath, santos2024assessing, dotch2023understanding, michel2024sounds}. However, our contribution lies in showing how such domain knowledge can be surfaced and operationalized through visualization. Rather than prescribing compositional rules, ResonaVis enables composers to discover and validate these relationships through data-driven exploration. From a VIS perspective, this demonstrates how visualization can bridge low-level acoustic features and high-level compositional intent, supporting insight generation in creative domains.

\subsection{RQ2: Interpretation and the Role of Interaction Data}

Our results further show that the value of ResonaVis emerges not only from individual visualizations, but from the integration of interaction data with audio features within a unified workflow (Section \ref{sec:perc}). Participants emphasized that interaction logs—capturing when and how layers were activated—provided essential context for interpreting musical properties. Without this temporal and behavioral grounding, acoustic features were perceived as static and less actionable.

This integration enables a shift from analyzing \textit{what} the music contains to understanding \textit{how} it unfolds through interaction. Visualizations such as Sankey diagrams and timelines supported tracing engagement patterns, identifying co-occurrence across layers, and reasoning about transitions over time. These findings align with prior work on visualization as a tool for exploratory analysis \cite{lima2021survey}, but extend it to a setting where interpretation is inherently creative and context-dependent.

Importantly, usability metrics (SUS scores of 70.71 and 73.75) indicate that the system is learnable and falls within the range typically considered acceptable in HCI studies \cite{bangor2008empirical}. However, usability alone does not fully capture the system’s impact. Participants also demonstrated consistent increases in confidence across multiple compositional and analytical dimensions (Section~\ref{sec:conf}), suggesting that ResonaVis supports not only interaction with the interface but also a deeper understanding of musical structure and data representations.

While the sample size is small n=7 (Study 1) and n=2 (Study 2), such sizes are common in visualization and HCI research, where studies emphasize depth of insight and formative evaluation over statistical generalization \cite{10.1145/169059.169166, faulkner2003beyond, lam2011empirical}. Prior work has also shown that parametric tests remain robust for small samples and Likert-scale data when effects are consistent, supporting the interpretation of observed confidence gains as indicative of meaningful trends \cite{norman2010likert, de2013using}. Accordingly, we interpret these results in conjunction with consistent directional improvements across participants.

Taken together, these findings suggest that, in creative workflows, the primary value of visualization lies in supporting interpretability and contextual reasoning. By linking interaction behavior with acoustic features, ResonaVis enables users to connect abstract data patterns with concrete compositional decisions, highlighting the importance of integrated, interpretable representations over isolated feature presentation.

\subsection{RQ3: Implications for Visualization in Creativity}

We demonstrate how visualization can support creative, interpretive tasks by linking ResonaVis design goals with observed user behavior:

\textbf{Support cross-modal reasoning.}
Aligned with our goal of integrating interaction and acoustic data (Section \ref{sec:cross model}), participants used combined views to relate engagement patterns to musical structure (e.g., sustained layer use with stable spectral energy), enabling deeper insight beyond isolated features (Sections \ref{sec:insights} and \ref{sec:qual}).

\textbf{Design for interpretability over completeness.}
Although the system provides feature-rich views, participants consistently favored interpretable representations, while more abstract features required effort but led to measurable confidence gains (Section~\ref{sec:conf}), reinforcing our emphasis on intuitive encodings.

\textbf{Scaffold domain-specific understanding.}
Consistent with our goal of supporting non-expert interpretation, participants initially struggled with abstract features (e.g., spectral properties) (Section \ref{sec:conf}) but demonstrated significant post-use confidence improvements (Section~\ref{sec:conf}), indicating that visualization can facilitate learning when grounded in domain context.

\textbf{Enable iterative exploration.}
Reflecting our exploratory workflow design, participants moved iteratively between inspection, pattern detection, and refinement, highlighting the importance of supporting non-linear analysis in creative tasks. These implications suggest that visualization systems for creative domains should prioritize integrated, interpretable, and workflow-oriented designs over isolated analytics.

\subsection{Limitations and Future Work}
This study has several limitations. \new{The uCue dataset used for visualization is modest, serving as formative design material rather than a representative sample.} \new{This evaluation focuses on early-stage ideation and exploratory analysis, not long-term compositional outcomes.} The visualizations represented only a subset of musical parameters, primarily rhythm, dynamics, spectral energy, and brightness, and may not encompass the full range of factors relevant to composition. \new{Finally, the participant pool is limited in size and diversity, which may restrict generalizability, and findings may reflect a narrow slice of compositional approaches. Relatedly, this study does not include clinical trials evaluating the therapeutic benefits, if any, of ResonaVis-composed music for children with ASD; our focus is on the visualization system's capacity to support composers' analytical and compositional reasoning, with clinical efficacy left to future work.} 

% This study has several limitations. The dataset from uCue, used for visualization is modest and serves as formative design material rather than a representative sample. This evaluation focuses on early-stage ideation and  exploratory analysis which  does not capture long-term compositional outcomes. The visualizations represented only a subset of musical parameters, primarily rhythm, dynamics, spectral energy, and brightness, and may not encompass the full range of factors relevant to composition.. Finally, the participant pool is limited in size and diversity, which may restrict generalizability, and findings may reflect may reflect a narrow slice of compositional approaches.

% While the visualizations are based on established techniques, their novelty lies in integration and application rather than new encoding design

Future work should \new{also} explore both technical and methodological extensions. On the technical side, incorporating adaptive components—such as pattern recognition or recommendation modules—could support more proactive guidance for insight generation. \new{At larger data scales, aggregation and filtering mechanisms will be needed to maintain legibility across encodings; child-level comparison across sessions is a natural extension of the current session-level design.} Developing alternative visual encodings that better represent simultaneity and hierarchical musical structure remains an open direction. On the methodological side, longitudinal studies are needed to understand how visualization supports full composition workflows over time. Expanding participation to include professional composers and therapists will further ground the system in real-world contexts. 

\section{Conclusion}
ResonaVis demonstrates the potential of visualization-driven approaches to support composers in reasoning about music for therapeutic contexts by making key elements—such as rhythm, dynamics, pitch, spectral energy, and brightness—more interpretable. Through interactive visualizations, the system surfaced patterns that enabled participants to reflect on compositional decisions, including aligning rhythmic and dynamic structures, balancing spectral properties, and designing smoother transitions to achieve cohesive musical outcomes. While these observations are consistent with prior work on therapeutic music, our contribution lies in showing how such relationships can be revealed and operationalized through visual analytics. More broadly, this work highlights how integrating interaction and audio data can support composer-centered workflows that move beyond passive analysis toward active exploration and decision-making. Future work will extend this approach by incorporating richer analytical methods, refining visualization designs, and exploring additional musical dimensions. 

% In doing so, ResonaVis opens new directions for leveraging visualization to support creative reasoning in complex, data-rich domains.

%\newpage
\section*{Acknowledgments}
% We utilized ChatGPT, Grammarly, and Overleaf’s autocorrect to assist with refining the text, improving grammar and clarity, and suggesting improvements to the paper’s flow, organization, and formatting. All outputs were reviewed and revised by the authors, who take full responsibility for the final content.
\add{This work was supported by the National Science Foundation via the Accelerating Research Translation Program, Grant No. 2331440, and UD Institute for Engineering-Driven Health. We would like to thank Connor Fone and Justin Eichenberg for co-designing the dashboard with us, Liam Stapley for revamping the UI, Dr. Kathleen McCoy for her advice, and Sensify Lab members for their feedback on the manuscript. ChatGPT, Grammarly, and Overleaf’s autocorrect were used to improve grammar, clarity, organization, and formatting. All suggestions were reviewed and revised by the authors, who take full responsibility for the final content.}
% \clearpage

\section*{\new{Supplemental Materials}}
\new{Our Supplementary Material contains study design for both the studies, the visualizations and their descriptions, and conversation prompts which were used to facilitate better conversation with the participants. They are available at \url{https://osf.io/xg6su}.}

\bibliographystyle{abbrv-doi-hyperref}

\bibliography{template}

@misc{mauriello2025interactive,
  title={Interactive musical interface to enhance formative listening experiences for children with autism spectrum disorder},
  author={Mauriello, Matthew and Stevens, Daniel and Ruggiero, Elise and Karwankar, Abhishek},
  year={2025},
  month=aug # "~21",
  publisher={Google Patents},
  note={US Patent App. 19/201,395},
  doi = {https://patents.google.com/patent/US20250265035A1/en}
}

@article{lam2011empirical,
  author={Lam, Heidi and Bertini, Enrico and Isenberg, Petra and Plaisant, Catherine and Carpendale, Sheelagh},
  journal={IEEE Trans. Visual Comput. Graphics}, 
  title={Empirical Studies in Information Visualization: Seven Scenarios}, 
  year={2012},
  volume={18},
  number={9},
  pages={1520-1536},
  doi={10.1109/TVCG.2011.279}
}

@article{norman2010likert,
  title={Likert scales, levels of measurement and the “laws” of statistics},
  author={Norman, Geoff},
  journal={Advances in health sciences education},
  volume={15},
  number={5},
  pages={625--632},
  year={2010},
  publisher={Springer},
  doi = {10.1007/s10459-010-9222-y}
}

@article{de2013using,
  title={Using the Student's" t"-Test with Extremely Small Sample Sizes.},
  author={De Winter, Joost CF},
  journal={Practical assessment, research \& evaluation},
  volume={18},
  number={10},
  pages={n10},
  year={2013},
  publisher={ERIC},
  doi = {https://doi.org/10.7275/e4r6-dj05}
}

@inproceedings{10.1145/169059.169166,
author = {Nielsen, Jakob and Landauer, Thomas K.},
title = {A mathematical model of the finding of usability problems},
year = {1993},
isbn = {0897915755},
publisher = {ACM},
address = {New York},
url = {https://doi.org/10.1145/169059.169166},
doi = {10.1145/169059.169166},
booktitle = {Proceedings of the INTERACT '93 and CHI '93 Conference on Human Factors in Computing Systems},
pages = {206–213},
numpages = {8},
location = {Amsterdam, The Netherlands},
series = {CHI '93}
}

@article{faulkner2003beyond,
  title={Beyond the five-user assumption: Benefits of increased sample sizes in usability testing},
  author={Faulkner, Laura},
  journal={Behavior Research Methods, Instruments, \& Computers},
  volume={35},
  number={3},
  pages={379--383},
  year={2003},
  publisher={Springer},
  doi = {https://doi.org/10.3758/BF03195514}
}

@incollection{wilkie2013towards,
  title={Towards a participatory approach for interaction design based on conceptual metaphor theory: a case study from music interaction},
  author={Wilkie, Katie and Holland, Simon and Mulholland, Paul},
  booktitle={Music and Human-Computer Interaction},
  pages={259--270},
  year={2013},
  publisher={Springer},
  doi = {https://doi.org/10.1007/978-1-4471-2990-5_15}
}

@article{grond2019participatory,
  author={Grond, Florian and Motta-Ochoa, Rossio and Miyake, Natalie and Tembeck, Tamar and Park, Melissa and Blain-Moraes, Stefanie},
  journal={IEEE Transactions on Affective Computing}, 
  title={Participatory Design of Affective Technology: Interfacing Biomusic and Autism}, 
  year={2022},
  volume={13},
  number={1},
  pages={250-261},
  doi={10.1109/TAFFC.2019.2922911}}

@article{clarke2017thematic,
  title={Thematic analysis},
  author={Clarke, Victoria and Braun, Virginia},
  journal={The journal of positive psychology},
  volume={12},
  number={3},
  pages={297--298},
  year={2017},
  publisher={Taylor \& Francis},
  doi = {10.1080/17439760.2016.1262613}
}

@article{sarikaya2018we,
  author={Sarikaya, Alper and Correll, Michael and Bartram, Lyn and Tory, Melanie and Fisher, Danyel},
  journal={IEEE Trans. Visual Comput. Graphics}, 
  title={What Do We Talk About When We Talk About Dashboards?}, 
  year={2019},
  volume={25},
  number={1},
  pages={682-692},
  doi={10.1109/TVCG.2018.2864903}
}

@incollection{keim2008visual,
  title={Visual analytics: Definition, process, and challenges},
  author={Keim, Daniel and Andrienko, Gennady and Fekete, Jean-Daniel and G{\"o}rg, Carsten and Kohlhammer, J{\"o}rn and Melan{\c{c}}on, Guy},
  booktitle={Information visualization: Human-centered issues and perspectives},
  pages={154--175},
  year={2008},
  publisher={Springer},
  doi = {https://doi.org/10.1007/978-3-540-70956-5_7}
}

@inproceedings{munzner2025visualization,
author = {Munzner, Tamara},
title = {Visualization Analysis and Design},
year = {2025},
isbn = {9798400715433},
publisher = {ACM},
address = {New York},
url = {https://doi.org/10.1145/3721241.3733989},
doi = {10.1145/3721241.3733989},
booktitle = {Proceedings of the SIGGRAPH Courses '25},
articleno = {10},
numpages = {2},
location = {
},

}

@article{sedlmair2012design,
  author={Sedlmair, Michael and Meyer, Miriah and Munzner, Tamara},
  journal={IEEE Trans. Visual Comput. Graphics}, 
  title={Design Study Methodology: Reflections from the Trenches and the Stacks}, 
  year={2012},
  volume={18},
  number={12},
  pages={2431-2440},
  doi={10.1109/TVCG.2012.213}
  }

@article{chang2009defining,
  title={Defining insight for visual analytics},
  author={Chang, Remco and Ziemkiewicz, Caroline and Green, Tera Marie and Ribarsky, William},
  journal={IEEE Computer Graphics and Applications},
  volume={29},
  number={2},
  pages={14--17},
  year={2009},
  publisher={IEEE}, 
  doi = {10.1109/MCG.2009.22}
}

@article{north2006toward,
  author={North, C.},
  journal={IEEE Computer Graphics and Applications}, 
  title={Toward measuring visualization insight}, 
  year={2006},
  volume={26},
  number={3},
  pages={6-9},
  doi={10.1109/MCG.2006.70}}

@inproceedings{pirolli2005sensemaking,
  title={The sensemaking process and leverage points for analyst technology as identified through cognitive task analysis},
  author={Pirolli, Peter and Card, Stuart},
  booktitle={Proceedings of international conference on intelligence analysis},
  volume={5},
  number={1},
  pages={2--4},
  year={2005},
  organization={McLean, VA, USA}
}

@ARTICLE{11262783,
  author={Xing, Yiwen and Ortoleva, Maria Teresa and Borgo, Rita and Abdul-Rahman, Alfie},
  journal={IEEE Trans. Visual Comput. Graphics}, 
  title={Collaborating Across Domains and Roles: An Interview Study of Visualization Design Practices}, 
  year={2026},
  volume={32},
  number={1},
  pages={571-581},
  doi={10.1109/TVCG.2025.3634711}}

@ARTICLE{10453452,
  author={Han, Yi and Abowd, Gregory D. and Stasko, John},
  journal={IEEE Trans. Visual Comput. Graphics}, 
  title={IntiVisor: A Visual Analytics System for Interaction Log Analysis}, 
  year={2025},
  volume={31},
  number={3},
  pages={1772-1784},
  doi={10.1109/TVCG.2024.3370637}}

@inproceedings{10.1145/3240925.3240958,
    author = {Cibrian, Franceli L. and Mercado, Jose and Escobedo, Lizbeth and Tentori, Monica},
    title = {A Step towards Identifying the Sound Preferences of Children with Autism},
    year = {2018},
    isbn = {9781450364508},
    publisher = {ACM},
    address = {New York},
    url = {https://doi.org/10.1145/3240925.3240958},
    doi = {10.1145/3240925.3240958},
    booktitle = {Proceedings of the 12th EAI International Conference on Pervasive Computing Technologies for Healthcare},
    pages = {158–167},
    numpages = {10},
    location = {New York, NY, USA},
    series = {PervasiveHealth '18}
}

@article{mayer2021music,
  title={Music therapy for children with autistic spectrum disorder and/or other neurodevelopmental disorders: a systematic review},
  author={Mayer-Benarous, Hanna and Benarous, Xavier and Vonthron, Fran{\c{c}}ois and Cohen, David},
  journal={Frontiers in psychiatry},
  volume={12},
  pages={435},
  year={2021},
  publisher={Frontiers},
  doi = {10.3389/fpsyt.2021.643234}
}

@article{geretsegger2014music,
  title={Music therapy for people with autism spectrum disorder},
  author={Geretsegger, Monika and Elefant, Cochavit and M{\"o}ssler, Karin A and Gold, Christian},
  journal={Cochrane Database of Systematic Reviews},
  number={6},
  year={2014},
  publisher={John Wiley \& Sons, Ltd},
  doi = {https://doi.org/10.1002/14651858.CD004381.pub3}
}

@article{kim2008effects,
  title={The effects of improvisational music therapy on joint attention behaviors in autistic children: a randomized controlled study},
  author={Kim, Jinah and Wigram, Tony and Gold, Christian},
  journal={Journal of autism and developmental disorders},
  volume={38},
  pages={1758--1766},
  year={2008},
  publisher={Springer},
  doi = {10.1007/s10803-008-0566-6}
}

@article{reschke2011history,
  title={History of music therapy treatment interventions for children with autism},
  author={Reschke-Hern{\'a}ndez, Alaine E},
  journal={Journal of Music Therapy},
  volume={48},
  number={2},
  pages={169--207},
  year={2011},
  publisher={American Music Therapy Association},
  doi = {10.1093/jmt/48.2.169}
}

@article{gold2004effects,
  title={Effects of music therapy for children and adolescents with psychopathology: a meta-analysis},
  author={Gold, Christian and Voracek, Martin and Wigram, Tony},
  journal={Journal of Child Psychology and Psychiatry},
  volume={45},
  number={6},
  pages={1054--1063},
  year={2004},
  publisher={Wiley Online Library},
  doi = {10.1111/j.1469-7610.2004.t01-1-00298.x}
}

@article{lagasse2014effects,
  title={Effects of a music therapy group intervention on enhancing social skills in children with autism},
  author={LaGasse, A Blythe},
  journal={Journal of music therapy},
  volume={51},
  number={3},
  pages={250--275},
  year={2014},
  publisher={Oxford University Press US},
  doi = {10.1093/jmt/thu012}
}

@article{lima2021survey,
author = {Lima, Hugo B. and Santos, Carlos G. R. Dos and Meiguins, Bianchi S.},
title = {A Survey of Music Visualization Techniques},
year = {2021},
issue_date = {September 2022},
publisher = {ACM},
address = {New York},
volume = {54},
number = {7},
issn = {0360-0300},
url = {https://doi.org/10.1145/3461835},
doi = {10.1145/3461835},
journal = {ACM Comput. Surv.},
month = jul,
articleno = {143},
numpages = {29}
}

@article{greene1987stakeholder,
title = {Stakeholder participation in evaluation design: Is it worth the effort?},
journal = {Evaluation and Program Planning},
volume = {10},
number = {4},
pages = {379-394},
year = {1987},
issn = {0149-7189},
doi = {https://doi.org/10.1016/0149-7189(87)90010-3},
url = {https://www.sciencedirect.com/science/article/pii/0149718987900103},
author = {Jennifer C. Greene}
}

@article{muller1993participatory,
author = {Muller, Michael J. and Kuhn, Sarah},
title = {Participatory design},
year = {1993},
issue_date = {June 1993},
publisher = {ACM},
address = {New York},
volume = {36},
number = {6},
issn = {0001-0782},
url = {https://doi.org/10.1145/153571.255960},
doi = {10.1145/153571.255960},
journal = {Commun. ACM},
month = jun,
pages = {24–28},
numpages = {5}
}

@article{henry1996creative,
  title={Creative processes in children's musical compositions: A review of the literature},
  author={Henry, Warren},
  journal={Update: Applications of Research in Music Education},
  volume={15},
  number={1},
  pages={10--15},
  year={1996},
  publisher={SAGE Publications Sage CA: Los Angeles, CA},
  doi = {https://doi.org/10.1177/87551233960150010}
}

@article{wigram2006music,
  title={Music therapy in the assessment and treatment of autistic spectrum disorder: clinical application and research evidence},
  author={Wigram, Tony and Gold, Christian},
  journal={Child: care, health and development},
  volume={32},
  number={5},
  pages={535--542},
  year={2006},
  publisher={Wiley Online Library},
  doi = {10.1111/j.1365-2214.2006.00615.x}
}

@article{whipple2004music,
  title={Music in intervention for children and adolescents with autism: A meta-analysis},
  author={Whipple, Jennifer},
  journal={Journal of music therapy},
  volume={41},
  number={2},
  pages={90--106},
  year={2004},
  publisher={American Music Therapy Association},
  doi = {10.1093/jmt/41.2.90}
}

@article{bharathi2019music,
  title={Music therapy as a therapeutic tool in improving the social skills of autistic children},
  author={Bharathi, Geetha and Venugopal, Anila and Vellingiri, Balachandar},
  journal={The Egyptian Journal of Neurology, Psychiatry and Neurosurgery},
  volume={55},
  pages={1--6},
  year={2019},
  publisher={Springer},
  doi = {https://doi.org/10.1186/s41983-019-0091-x}
}

@article{lagasse2017social,
  title={Social outcomes in children with autism spectrum disorder: a review of music therapy outcomes},
  author={LaGasse, A Blythe},
  journal={Patient related outcome measures},
  pages={23--32},
  year={2017},
  publisher={Taylor \& Francis},
  doi = {10.2147/PROM.S106267}
}

@article{chen2024music,
  title={Music-based therapeutic interventions for medical school students with emotional regulation and mental health: a pre-post cohort study},
  author={Chen, Quan and Mao, Chaoqin and Qi, Laihua and Luo, Yang and Yang, Guangyao and Wang, Lei and Liu, Chen and Zheng, Chuansheng and Zhang, Jinxiang and Fan, Cheng},
  journal={Frontiers in Psychology},
  volume={15},
  pages={1401129},
  year={2024},
  publisher={Frontiers Media SA},
  doi = {10.3389/fpsyg.2024.1401129}
}

@article{katagiri2009effect,
  title={The effect of background music and song texts on the emotional understanding of children with autism},
  author={Katagiri, June},
  journal={Journal of music therapy},
  volume={46},
  number={1},
  pages={15--31},
  year={2009},
  publisher={American Music Therapy Association},
  doi = {10.1093/jmt/46.1.15}
}

@article{de2017understanding,
  title={Understanding the structure of musical compositions: Is visualization an effective approach?},
  author={De Prisco, Roberto and Malandrino, Delfina and Pirozzi, Donato and Zaccagnino, Gianluca and Zaccagnino, Rocco},
  journal={Information Visualization},
  volume={16},
  number={2},
  pages={139--152},
  year={2017},
  publisher={SAGE Publications Sage UK: London, England},
  doi = {10.1177/1473871616655468}
}

@article{bernard2015visual,
  title={A visual-interactive system for prostate cancer cohort analysis},
  author={Bernard, J{\"u}rgen and Sessler, David and May, Thorsten and Schlomm, Thorsten and Pehrke, Dirk and Kohlhammer, J{\"o}rn},
  journal={IEEE computer graphics and applications},
  volume={35},
  number={3},
  pages={44--55},
  year={2015},
  publisher={IEEE},
  doi = {10.1109/MCG.2015.49}
}

@article{naranjo2019visual,
  title={A visual dashboard to track learning analytics for educational cloud computing},
  author={Naranjo, Diana M and Prieto, Jos{\'e} R and Molt{\'o}, Germ{\'a}n and Calatrava, Amanda},
  journal={Sensors},
  volume={19},
  number={13},
  pages={2952},
  year={2019},
  publisher={MDPI},
  doi = {https://doi.org/10.3390/s19132952}
}

@article{franklin2017dashboard,
title = {Dashboard visualizations: Supporting real-time throughput decision-making},
journal = {Journal of Biomedical Informatics},
volume = {71},
pages = {211-221},
year = {2017},
issn = {1532-0464},
doi = {https://doi.org/10.1016/j.jbi.2017.05.024},
url = {https://www.sciencedirect.com/science/article/pii/S1532046417301235},
author = {Amy Franklin and Swaroop Gantela and Salsawit Shifarraw and Todd R. Johnson and David J. Robinson and Brent R. King and Amit M. Mehta and Charles L. Maddow and Nathan R. Hoot and Vickie Nguyen and Adriana Rubio and Jiajie Zhang and Nnaemeka G. Okafor}
}

@article{cybulski2015creative,
title = {Creative problem solving in digital space using visual analytics},
journal = {Computers in Human Behavior},
volume = {42},
pages = {20-35},
year = {2015},
note = {Digital Creativity: New Frontier for Research and Practice},
issn = {0747-5632},
doi = {https://doi.org/10.1016/j.chb.2013.10.061},
url = {https://www.sciencedirect.com/science/article/pii/S0747563213004111},
author = {Jacob L. Cybulski and Susan Keller and Lemai Nguyen and Dilal Saundage}
}

@article{mcadams2013musical,
  title={Musical timbre perception},
  author={McAdams, Stephen},
  journal={The psychology of music},
  volume={3},
  year={2013},
  doi = {10.1016/B978-0-12-381460-9.00002-X}
}

@inproceedings{hiraga2002performance,
author = {Hiraga, Rumi and Mizaki, Reiko and Fujishiro, Issei},
title = {Performance visualization: a new challenge to music through visualization},
year = {2002},
isbn = {158113620X},
publisher = {ACM},
address = {New York},
url = {https://doi.org/10.1145/641007.641054},
doi = {10.1145/641007.641054},
booktitle = {Proceedings of the Tenth ACM MULTIMEDIA '02},
pages = {239–242},
numpages = {4},
location = {Juan-les-Pins, France}
}

@article{fonteles2013creating,
title = {Creating and evaluating a particle system for music visualization},
journal = {Journal of Visual Languages and Computing},
volume = {24},
number = {6},
pages = {472-482},
year = {2013},
issn = {1045-926X},
doi = {https://doi.org/10.1016/j.jvlc.2013.10.002},
url = {https://www.sciencedirect.com/science/article/pii/S1045926X13000566},
author = {Joyce Horn Fonteles and Maria Andréia Formico Rodrigues and Victor Emanuel Dias Basso}
}

@article{born2005musical,
  title={On musical mediation: Ontology, technology and creativity},
  author={Born, Georgina},
  journal={Twentieth-century music},
  volume={2},
  number={1},
  pages={7--36},
  year={2005},
  publisher={Cambridge University Press},
  doi = {10.1017/S147857220500023X}
}

@article{de2012creativity,
  title={Creativity-centred design for ubiquitous musical activities: Two case studies},
  author={De Lima, Maria Helena and Keller, Dami{\'a}n and Pimenta, Marcelo Soares and Lazzarini, Victor and Miletto, Evandro Manara},
  journal={Journal of Music, Technology and Education},
  volume={5},
  number={2},
  pages={195--222},
  year={2012},
  publisher={Intellect},
  doi = {https://doi.org/10.1386/jmte.5.2.195_1}
}

@inproceedings{miller2022augmenting,
  title={Augmenting digital sheet music through visual analytics},
  author={Miller, Matthias and F{\"u}rst, Daniel and Hauptmann, Hanna and Keim, Daniel A and El-Assady, Mennatallah},
  booktitle={Computer Graphics Forum},
  volume={41},
  number={1},
  pages={301--316},
  year={2022},
  organization={Wiley Online Library},
  doi = {10.1111/cgf.14436}
}

@article{chan2009visualizing,
  title={Visualizing the semantic structure in classical music works},
  author={Chan, Wing-Yi and Qu, Huamin and Mak, Wai-Ho},
  journal={IEEE Trans. Visual Comput. Graphics},
  volume={16},
  number={1},
  pages={161--173},
  year={2009},
  publisher={IEEE},
  doi = {10.1109/TVCG.2009.63}
}

@book{smith2009practice,
  title={Practice-led research, research-led practice in the creative arts},
  author={Smith, Hazel},
  year={2009},
  publisher={Edinburgh University Press},
  doi = {https://www.jstor.org/stable/10.3366/j.ctt1g0b594}
}

@inproceedings{bjork2005games,
  title={Game design patterns},
  author={Bj{\"o}rk, Staffan and Lundgren, Sus and Holopainen, Jussi},
  booktitle={Proceedings of DiGRA 2003 Conference: Level Up},
  year={2003},
  doi = {https://doi.org/10.26503/dl.v2003i1.60}
}

@book{collins2010introduction,
  title={Introduction to computer music},
  author={Collins, Nick},
  year={2010},
  publisher={John Wiley \& Sons},
  doi = {10.1017/S1355771810000221}
}

@article{paine2009towards,
  title={Towards unified design guidelines for new interfaces for musical expression},
  author={Paine, Garth},
  journal={Organised Sound},
  volume={14},
  number={2},
  pages={142--155},
  year={2009},
  publisher={Cambridge University Press},
  doi = {10.1017/S1355771809000259}
}

@article{monroe2013temporal,
  author={Monroe, Megan and Lan, Rongjian and Lee, Hanseung and Plaisant, Catherine and Shneiderman, Ben},
  journal={IEEE Trans. Visual Comput. Graphics}, 
  title={Temporal Event Sequence Simplification}, 
  year={2013},
  volume={19},
  number={12},
  pages={2227-2236},
  doi={10.1109/TVCG.2013.200}
}

@inproceedings{mauriello2016simplifying,
author = {Mauriello, Matthew Louis and Shneiderman, Ben and Du, Fan and Malik, Sana and Plaisant, Catherine},
title = {Simplifying Overviews of Temporal Event Sequences},
year = {2016},
isbn = {9781450340823},
publisher = {ACM},
address = {New York},
url = {https://doi.org/10.1145/2851581.2892440},
doi = {10.1145/2851581.2892440},
booktitle = {Proceedings of the 2016 CHI EA on Human Factors in Computing Systems},
pages = {2217–2224},
numpages = {8},
location = {San Jose, California, USA},
series = {CHI EA '16}
}

@article{allen2009hath,
  title={Hath charms to soothe...' An exploratory study of how high-functioning adults with ASD experience music},
  author={Allen, Rory and Hill, Elizabeth and Heaton, Pam},
  journal={Autism},
  volume={13},
  number={1},
  pages={21--41},
  year={2009},
  publisher={Sage Publications Sage UK: London, England},
  doi = {10.1177/1362361307098511}
}

@inproceedings{dotch2023understanding,
author = {Hicks, Emani and Johnson, Jazette and Black, Rebecca W. and Hayes, Gillian R},
title = {Understanding Noise Sensitivity through Interactions in Two Online Autism Forums},
year = {2023},
isbn = {9798400702204},
publisher = {ACM},
address = {New York},
url = {https://doi.org/10.1145/3597638.3608413},
doi = {10.1145/3597638.3608413},
booktitle = {Proceedings of the ACM SIGACCESS ASSETS '23},
articleno = {19},
numpages = {12},
location = {New York, NY, USA}

}

@article{michel2024sounds,
  title={Sounds Pleasantness Ratings in Autism: Interaction Between Social Information and Acoustical Noise Level},
  author={Michel, Lisa and Ricou, Camille and Bonnet-Brilhault, Fr{\'e}d{\'e}rique and Houy-Durand, Emannuelle and Latinus, Marianne},
  journal={Journal of Autism and Developmental Disorders},
  volume={54},
  number={6},
  pages={2148--2157},
  year={2024},
  publisher={Springer},
  doi = {10.1007/s10803-023-05989-6}
}

@article{santos2024assessing,
  title={ASSESSING MUSICAL PREFERENCES OF CHILDREN ON THE AUTISTIC SPECTRUM: IMPLICATIONS FOR THERAPY},
  author={SANTOS, Nat{\'a}lia and BERNARDES, Gilberto and COTTA, Rita and COELHO, Nilzabeth and BAGANHA, Alessandra},
  year={2024},
  doi = {10.13140/RG.2.2.17204.67207}
}

@inproceedings{hart2006nasa,
  title={NASA-task load index (NASA-TLX); 20 years later},
  author={Hart, Sandra G},
  booktitle={Proceedings of the human factors and ergonomics society annual meeting},
  volume={50},
  number={9},
  pages={904--908},
  year={2006},
  organization={Sage publications Sage CA: Los Angeles, CA},
  doi = {https://doi.org/10.1177/154193120605000909}
}

@article{vlachogianni2022perceived,
  title={Perceived usability evaluation of educational technology using the System Usability Scale (SUS): A systematic review},
  author={Vlachogianni, Prokopia and Tselios, Nikolaos},
  journal={Journal of Research on Technology in Education},
  volume={54},
  number={3},
  pages={392--409},
  year={2022},
  publisher={Taylor \& Francis},
  doi = {https://doi.org/10.1080/15391523.2020.1867938}
}

@book{sawyer2024explaining,
  title={Explaining creativity: The science of human innovation},
  author={Sawyer, Robert Keith and Henriksen, Danah},
  year={2024},
  publisher={Oxford university press},
  doi = {https://doi.org/10.1093/oso/9780197747537.001.0001}
}

@article{hargreaves2018musical,
  title={Musical identities mediate musical development},
  author={Hargreaves, David J and MacDonald, Raymond and Miell, Dorothy},
  journal={Music and music education in people’s lives},
  volume={1},
  pages={124--142},
  year={2018},
  publisher={Oxford University Press},
  doi = {https://doi.org/10.1093/oxfordhb/9780199730810.013.0008_update_001}
}

@article{mcferran2014community,
  title={Community music therapy in schools: Realigning with the needs of contemporary students, staff and systems},
  author={McFerran, Katrina Skewes and Rickson, Daphne},
  journal={International Journal of Community Music},
  volume={7},
  number={1},
  pages={75--92},
  year={2014},
  publisher={Intellect},
  doi = {https://doi.org/10.1386/ijcm.7.1.75_1}
}

@article{shneiderman2004designing,
  author = {Shneiderman, Ben},
title = {Designing for fun: how can we design user interfaces to be more fun?},
year = {2004},
issue_date = {September + October 2004},
publisher = {Association for Computing Machinery},
address = {New York, NY, USA},
volume = {11},
number = {5},
issn = {1072-5520},
url = {https://doi.org/10.1145/1015530.1015552},
doi = {10.1145/1015530.1015552},
journal = {Interactions},
month = sep,
pages = {48–50},
numpages = {3}
}

@article{bangor2008empirical,
  title={An empirical evaluation of the system usability scale},
  author={Bangor, Aaron and Kortum, Philip T and Miller, James T},
  journal={Intl. Journal of Human--Computer Interaction},
  volume={24},
  number={6},
  pages={574--594},
  year={2008},
  publisher={Taylor \& Francis},
  doi = {10.1080/10447310802205776}
}

@inproceedings{10.1145/964696.964700,
author = {Goto, Masataka},
title = {SmartMusicKIOSK: music listening station with chorus-search function},
year = {2003},
isbn = {1581136366},
publisher = {ACM},
address = {New York},
url = {https://doi.org/10.1145/964696.964700},
doi = {10.1145/964696.964700},
booktitle = {Proceedings of the 16th Annual ACM UIST '03},
pages = {31–40},
numpages = {10},
location = {Vancouver, Canada}

}

@inproceedings{10.1145/964696.964698,
author = {Begole, James "Bo" and Tang, John C. and Hill, Rosco},
title = {Rhythm modeling, visualizations and applications},
year = {2003},
isbn = {1581136366},
publisher = {ACM},
address = {New York},
url = {https://doi.org/10.1145/964696.964698},
doi = {10.1145/964696.964698},
booktitle = {Proceedings of the 16th Annual ACM UIST '03},
pages = {11–20},
numpages = {10},
location = {Vancouver, Canada}

}

@inbook{10.1145/3713043.3727053,
author = {Karwankar, Abhishek and Ruggiero, Elise and Lipkin, Zoe and Iyer, Malika Karthik and Brugel, Simon and Khatiwada, Prerana and Stevens, Daniel and Mauriello, Matthew Louis},
title = {uCue: An Interactive Musical Interface to Enhance Formative Listening Experiences for Children with ASD},
year = {2025},
isbn = {9798400714733},
publisher = {ACM IDC '25},
address = {New York},
doi = {10.1145/3713043.3727053},
url = {https://doi.org/10.1145/3713043.3727053},
booktitle = {Proceedings of the 24th Interaction Design and Children},
pages = {340–357},
numpages = {18}
}

@inproceedings{tsiris2020impact,
  title={The Impact Areas Questionnaire (IAQ): a music therapy service evaluation tool},
  author={Tsiris, Giorgos and Spiro, Neta and Coggins, Owen and Zubala, Ania},
  booktitle={Voices: A World Forum for Music Therapy},
  volume={20},
  number={2},
  year={2020},
  organization={GAMUT-Grieg Academy Music Therapy Research Centre (NORCE \& University of Bergen)},
  doi = {https://doi.org/10.15845/voices.v20i2.2816}
}

@article{daniel2022rhythmic,
  title={Rhythmic relating: Bidirectional support for social timing in autism therapies},
  author={Daniel, Stuart and Wimpory, Dawn and Delafield-Butt, Jonathan T and Malloch, Stephen and Holck, Ulla and Geretsegger, Monika and Tortora, Suzi and Osborne, Nigel and Sch{\"o}gler, Benjaman and Koch, Sabine and others},
  journal={Frontiers in Psychology},
  volume={13},
  pages={793258},
  year={2022},
  publisher={Frontiers Media SA},
  doi = {10.3389/fpsyg.2022.793258}
}

@article{khyzhna2020music,
  title={Music therapy as an important element in shaping communication competences in children with autism spectrum disorder},
  author={Khyzhna, Olga and Shafranska, Karina},
  journal={Journal of History Culture and Art Research},
  volume={9},
  number={3},
  pages={106--114},
  year={2020},
  doi = {10.7596/taksad.v9i3.2823}
}

@inproceedings{10.5555/832277.834354,
  author={Shneiderman, B.},
  booktitle={Proceedings 1996 IEEE Symposium on Visual Languages}, 
  title={The eyes have it: a task by data type taxonomy for information visualizations}, 
  year={1996},
  volume={},
  number={},
  pages={336-343},
  doi={10.1109/VL.1996.545307}}

@inproceedings{foote1999visualizing,
author = {Foote, Jonathan},
title = {Visualizing music and audio using self-similarity},
year = {1999},
isbn = {1581131518},
publisher = {ACM},
address = {New York},
url = {https://doi.org/10.1145/319463.319472},
doi = {10.1145/319463.319472},
booktitle = {Proceedings of the Seventh ACM MULTIMEDIA '99},
pages = {77–80},
numpages = {4},
location = {Orlando, Florida, USA}
}

@inproceedings{astrid2017moment,
  title={In-the-moment and beyond: Combining post-hoc and real-time data for the study of audience perception of electronic music performance},
  author={Astrid Bin, SM and Morreale, Fabio and Bryan-Kinns, Nick and McPherson, Andrew P},
  booktitle={IFIP conference on human-computer interaction},
  pages={263--281},
  year={2017},
  organization={Springer},
  doi = {10.1007/978-3-319-67744-6_18}
}

@book{wigram2004improvisation,
  title={Improvisation: Methods and techniques for music therapy clinicians, educators, and students},
  author={Wigram, Tony},
  year={2004},
  publisher={Jessica Kingsley Publishers},
  doi = {https://books.google.com/books?id=k9EPBQAAQBAJ}
}

@inproceedings{fiebrink2011human,
author = {Fiebrink, Rebecca and Cook, Perry R. and Trueman, Dan},
title = {Human model evaluation in interactive supervised learning},
year = {2011},
isbn = {9781450302289},
publisher = {ACM},
address = {New York},
url = {https://doi.org/10.1145/1978942.1978965},
doi = {10.1145/1978942.1978965},
booktitle = {Proceedings of the SIGCHI Conference on Human Factors in Computing Systems},
pages = {147–156},
numpages = {10},
location = {Vancouver, BC, Canada},
series = {CHI '11}
}

@article{xambo2024human,
  title={Human--machine agencies in live coding for music performance},
  author={Xamb{\'o}, Anna and Roma, Gerard},
  journal={Journal of New Music Research},
  volume={53},
  number={1-2},
  pages={33--46},
  year={2024},
  publisher={Taylor \& Francis},
  doi = {https://doi.org/10.1080/09298215.2024.2442355}
}

@book{nielsen1994usability,
author = {Nielsen, Jakob},
title = {Usability Engineering},
year = {1994},
isbn = {9780080520292},
publisher = {Morgan Kaufmann Publishers Inc.},
address = {San Francisco, CA, USA},
doi = {https://dl.acm.org/doi/10.5555/2821575}
}

@book{lazar2017research,
  title={Research methods in human-computer interaction},
  author={Lazar, Jonathan and Feng, Jinjuan Heidi and Hochheiser, Harry},
  year={2017},
  publisher={Morgan Kaufmann},
  doi = {https://doi.org/10.1016/B978-0-12-805390-4.09987-8}
}

% \newpage
% \section{Appendix}
\appendix
% \vspace{-7mm}
\section{Appendix}
We include our positionality statement, design challenges in music composition, the visualizations in the dashboard, our qualitative codebook, and the qualitative feedback analysis in the appendix. 

\subsection{Positionality of the Research Team}
As a multidisciplinary team, we brought together expertise in music composition, data visualization, and experience working with children on the autism spectrum. This combination informed both the framing of the problem—understanding interaction-driven composition—and the selection of visualization techniques that balance analytical rigor with musical interpretability.

\FloatBarrier 
\Needspace{0.45\textheight}

\subsection{Design Challenges in Music Composition}
\label{app:challenges}

\begin{table}[H]
\caption{\textbf{Challenges in music composition and their descriptions.}}
\label{tab:Challenges_appendix}
\centering
\renewcommand{\arraystretch}{1.2}
\rowcolors{2}{gray!10}{white}
\small
\begin{tabularx}{\columnwidth}{>{\raggedright\arraybackslash}p{0.34\columnwidth} X}
\hline
\textbf{Challenge} & \textbf{Description} \\
\hline

\textbf{Finding Universally Appealing Sounds} &
Identifying sounds that resonate with a general audience, considering significant variability in listener preferences. \\

\textbf{Sound Compatibility} &
Determining which sounds work well together in a composition by exploring harmonic and rhythmic compatibility between musical elements. \\

\textbf{Balancing Complexity and Simplicity} &
Creating musically complex pieces that remain satisfying yet simple to the ear, balancing layering and dynamics. \\

\textbf{Generating New Musical Ideas} &
Supporting innovation in composition through tools that inspire and facilitate the creation of new ideas. \\

\textbf{Identifying Listener Preferences} &
Understanding and predicting listener preferences to craft compositions that are well received, using creative insight and empirical testing. \\

\textbf{Designing Listener Preference Tests} &
Developing effective methods to test and measure listener preferences to refine compositions and ensure appeal. \\

\textbf{Visualizing Sonic Space} &
Providing tools to visualize the sonic space each instrument occupies within the hearing range, aiding in balancing overall sound. \\

\textbf{Visualizing Track Relationships} &
Understanding the relationships between different tracks in a larger composition to create cohesive musical pieces. \\

\hline
\end{tabularx}
\end{table}

\FloatBarrier 
\Needspace{0.45\textheight}

\subsection{Dashboard Visualizations}
\label{app:plots}

\begin{table}[H]
\caption{\textbf{Dashboard visualizations, their descriptions, challenges addressed, and parameters visualized.}}
\label{tab:plots_appendix}
\centering
\renewcommand{\arraystretch}{1.15}
\setlength{\tabcolsep}{2.5pt}
\rowcolors{2}{gray!10}{white}
\footnotesize

\begin{tabularx}{\columnwidth}{
>{\raggedright\arraybackslash}p{0.20\columnwidth}
>{\raggedright\arraybackslash}X
>{\raggedright\arraybackslash}X
>{\raggedright\arraybackslash}p{0.20\columnwidth}
}
\hline
\textbf{Plot Name} & 
\textbf{Description} & 
\textbf{Addresses Challenge} & 
\textbf{Parameters Visualized} \\
\hline

\textbf{Sankey Diagram} &
Illustrates transitions between audio layers, showing the flow from one layer to another. &
Understanding relationships and transitions between layers. &
Layer transitions; interaction sequences \\

\textbf{Covariance Matrix} &
Shows relationships between layers and indicates frequently used combinations. &
Identifying compatible sound combinations. &
Layer compatibility; interaction frequency \\

\textbf{Bar Graph with Error Bars} &
Displays the most frequently used layers across participants, with error bars showing usage variability. &
Balancing complexity and simplicity based on listener preferences. &
Interaction frequency; usage variability \\

\textbf{Full Session Diagram} &
Provides a timeline of layer activations for each song, showing interaction patterns over time. &
Visualizing the temporal structure of compositions. &
Layer activations; temporal patterns \\

\textbf{Time Series Plot} &
Tracks interactions with audio layers across the session. &
Visualizing engagement and how tracks relate over time. &
Engagement levels; temporal interactions \\

\textbf{Waveplot} &
Visualizes changes in audio amplitude over time, representing energy dynamics. &
Balancing complexity and simplicity by visualizing dynamic range. &
Amplitude; loudness variations \\

\textbf{MFCC} &
Visualizes frequency-related audio features that approximate aspects of perceived timbre. &
Visualizing the sonic space occupied by instruments. &
Frequency components; timbral features \\

\textbf{Spectrogram} &
Displays frequency content over time, showing harmonic and melodic structure. &
Analyzing pitch and frequency relationships. &
Frequency; intensity; pitch dynamics \\

\textbf{Spectral Centroid} &
Shows the center of spectral energy, indicating sound brightness over time. &
Crafting complex yet satisfying pieces. &
Spectral energy; brightness characteristics \\

\textbf{Spectral Rolloff} &
Indicates where most spectral energy is concentrated, revealing sound characteristics. &
Manipulating frequency characteristics to match listener preferences. &
Spectral energy distribution; frequency concentration \\

\hline
\end{tabularx}
\end{table}

\FloatBarrier 
\Needspace{0.45\textheight}

\subsection{Qualitative Codebook}
\label{app:codebook}

\begin{table}[H]
\caption{\textbf{Codebook for qualitative analysis of participant responses.}}
\label{tab:codebook_appendix}
\centering
\renewcommand{\arraystretch}{1.15}
\setlength{\tabcolsep}{2pt}
\rowcolors{2}{gray!10}{white}
\scriptsize

\begin{tabularx}{\columnwidth}{
>{\raggedright\arraybackslash}p{0.19\columnwidth}
>{\raggedright\arraybackslash}p{0.21\columnwidth}
>{\raggedright\arraybackslash}X
>{\raggedright\arraybackslash}p{0.25\columnwidth}
}
\hline
\textbf{Theme} & 
\textbf{Code} & 
\textbf{Definition} & 
\textbf{Representative Example from Data} \\
\hline

\textbf{Rhythmic Alignment} &
Matching rhythmic spikes &
Layers that exhibit aligned rhythmic peaks are perceived as more cohesive. &
``Layers with matching or overlapping spikes sounded well together.'' \\

&
Rhythmic contrast &
Contrast between steady ambient sounds and broken melodies creates diversity. &
``The stochasticity of ambient sounds contrasted with a stable percussive bass line.'' \\

&
Rhythmic dominance &
Certain layers, particularly high-beat tracks, tend to dominate the song. &
``High beat tracks frequently dominated, while ambient noises remained steady.'' \\

\textbf{Dynamic Balance} &
Amplitude complementarity &
Layers with different amplitude ranges complement each other. &
``Layers with differing amplitude ranges complemented each other without clashing.'' \\

&
Dynamic hierarchy &
Certain layers stand out while others play a supporting role. &
``A hierarchy was established, where certain layers stood out prominently.'' \\

&
Clustered dynamic spikes &
Blended layers had energy peaks closer together, improving harmony. &
``Layers with spikes closer together blended more harmoniously.'' \\

\textbf{Spectral Energy \& Brightness} &
Low-frequency preference &
Participants preferred layers with lower frequencies, which may be calming. &
``Lower brightness levels were observed in the middle to low end of the graph.'' \\

&
Spectral energy consistency &
Layers that maintain stable spectral energy across the song were perceived as structured. &
``Harp spanned the spectrum but concentrated at lower frequencies.'' \\

&
Percussion dominance &
Percussion layers had higher frequencies and overtones, creating a strong auditory presence. &
``Percussion covered the entire frequency spectrum, often overshadowing the bass drum.'' \\

\textbf{Transitions \& Auditory Experience} &
Overtones contrast &
Transitions from complex overtones to simpler ones create a sense of contrast. &
``Children often switched from melody to harmony, providing a satisfying transition.'' \\

&
Frequency shifts &
Transitions commonly lowered frequency, creating a calming effect. &
``Frequency generally lowered as transitions occurred.'' \\

&
Percussion-driven variability &
Percussion layers increased variation, making transitions more dynamic. &
``Percussion enhanced the auditory experience of other layers by adding rhythmic variety.'' \\

\hline
\end{tabularx}
\end{table}

\FloatBarrier 
\Needspace{0.45\textheight}
\subsection{Qualitative Feedback Analysis}
\label{app:qual_feedback}

\begin{table}[H]
\caption{\textbf{Summary of participants' feedback on music composition before and after the session, grouped thematically according to the challenges.}}
\label{tab:qualitative_feedback_appendix}
\centering
\renewcommand{\arraystretch}{1.15}
\setlength{\tabcolsep}{2.5pt}
\rowcolors{2}{gray!10}{white}
\scriptsize

\begin{tabularx}{\columnwidth}{
>{\raggedright\arraybackslash}p{0.27\columnwidth}
>{\raggedright\arraybackslash}X
>{\raggedright\arraybackslash}X
}
\hline
\textbf{Theme} & 
\textbf{Pre-Session Observations} & 
\textbf{Post-Session Changes} \\
\hline

\textbf{Rhythmic and Dynamic Elements (T1)} &
Preferred simple, steady, and predictable rhythms; dynamic ranges were kept soft or moderate. P1 and P3 emphasized easy-to-follow rhythms. P4 and P6 emphasized consistency and avoiding complexity. P7 and P8 emphasized small dynamic ranges and steady rhythm. &
Participants considered more rhythmic variation after the session, particularly P1, P2, P4, P6, and P7. They also considered greater use of percussion and transitions. Some participants, such as P8, maintained simplicity but adjusted based on children's responses. \\

\textbf{Pitch, Range, Timbre, and Overtones (T2)} &
Participants were cautious with pitch selection, favoring mid-to-lower pitch ranges and avoiding overstimulating timbres. P1 and P7 preferred warmer timbres, while P2, P3, P4, and P6 emphasized lower or mid-range pitches. &
Participants showed greater openness to timbral variation, particularly P1, P4, and P6. Some retained a preference for mellow and consistent ranges, including P7 and P8. Others made minor pitch-range adjustments to enhance dynamics, including P2 and P3. \\

\textbf{Spectral Energy and Brightness of Sound (T3)} &
Participants generally preferred darker or moderately bright sounds, particularly P3, P6, P7, and P8. P1 and P4 were open to brighter sounds but expressed caution. &
Participants made greater efforts to balance spectral energy and brightness, particularly P1, P4, and P7. Some retained a darker sound preference, such as P6 and P8, while also considering more dynamic contrasts. \\

\textbf{Adaptive Strategies Post-Session} &
Initial strategies were more rigid and based primarily on prior knowledge. &
Participants adapted their compositions more dynamically based on observed responses and data insights. More variation in rhythm, pitch, and timbre was introduced. \\

\hline
\end{tabularx}
\end{table}

\end{document}